\documentclass[pre,onecolumn,showpacs,aps]{revtex4}

\usepackage{amsfonts}
\usepackage{amsmath}
\usepackage{amssymb}
\usepackage{graphicx}
\usepackage{dcolumn}
\usepackage{times}

\begin{document}

\title{\bf Multiple atomic dark solitons in cigar-shaped Bose-Einstein condensates \\}

\author{
G.\ Theocharis$^{1}$,
A.\ Weller$^{2}$,
J.P.\ Ronzheimer$^{2}$,
C.\ Gross$^{2}$,
M.K.\ Oberthaler$^{2}$,
P.G.\ Kevrekidis$^{1}$, and
D.J.\ Frantzeskakis$^{3}$}
\affiliation{
$^{1}$ Department of Mathematics and Statistics,University of Massachusetts, Amherst MA 01003-4515, USA \\
$^{2}$ Kirchhoff Institut f\"{u}r Physik, INF 227, Universit\"{a}t Heidelberg, 69120, Heidelberg, Germany \\
$^{3}$ Department of Physics, University of Athens, Panepistimiopolis,Zografos, Athens 157 84, Greece}

\begin{abstract}
We consider the stability and dynamics of multiple dark solitons in
cigar-shaped Bose-Einstein condensates (BECs). Our study is
motivated by the fact that multiple matter-wave dark solitons may
naturally form in such settings as per our recent work [Phys. Rev.
Lett. {\bf 101}, 130401 (2008)]. First, we study the dark soliton
interactions and show that the dynamics of well-separated solitons
(i.e., ones that undergo a collision with relatively low velocities)
can be analyzed by means of particle-like equations of motion. The
latter take into regard the repulsion between solitons (via an
effective repulsive potential) and the confinement and
dimensionality of the system (via an effective parabolic trap for
each soliton). Next, based on the fact that stationary,
well-separated dark multi-soliton states emerge as a nonlinear
continuation of the appropriate excited eigensates of the quantum
harmonic oscillator, we use a Bogoliubov-de Gennes analysis to
systematically study the stability of such structures. We find that
for a sufficiently large number of atoms, multiple soliton states
may be dynamically stable, while for a small number of atoms, we
predict a dynamical instability emerging from resonance effects
between the eigenfrequencies of the soliton modes and the intrinsic excitation frequencies
of the condensate. Finally we present experimental realizations of
multi-soliton states including a three-soliton state consisting of
two solitons oscillating around a stationary one.
\end{abstract}

\maketitle

\section{Introduction}

Dark solitons, namely localized density dips on top of a stable
continuous-wave background
(or a background of finite extent) with a phase jump across their density
minimum, are fundamental
envelope solitons supported in nonlinear dispersive media. These
nonlinear waves emerge in
media with a positive (negative) group-velocity dispersion and
defocusing (focusing) nonlinearity, with
a proper model describing their evolution being the so-called
defocusing nonlinear Schr\"{o}dinger (NLS)
equation \cite{zsd}. Dark solitons have been studied extensively in the field of nonlinear optics
(from which the term ``dark'' was coined) \cite{kivpr}, but they have also
been observed in diverse physical
contexts, including liquids \cite{den1}, mechanical systems \cite{den2},
magnetic films \cite{magn}, and so on.

More recently, dark solitons have attracted much attention in the physics of Bose-Einstein condensates
(BECs) \cite{pethick,book2}, where they appear as fundamental macroscopic
nonlinear excitations of BECs with
repulsive interatomic interactions \cite{BECBOOK,revnonlin}.
It is, therefore, not surprising that experimental
work on {\it matter-wave dark solitons} started as early as ten years
ago \cite{han1,nist,dutton,bpa,han2}
and, very recently, continued even more
intensively \cite{engels,hamburg,hambcol,kip,technion}.
These important ``new age'' experiments highlighted various salient
features of dark solitons, verified previous
theoretical predictions and offered motivation for
further investigations.
A pertinent example is the creation of more than one matter-wave dark
solitons \cite{hambcol,kip}
(see also Ref. \cite{engels}) in cigar-shaped condensates, which were allowed to interact.
This invites a revisiting of the topic of dark solitons and especially
of their interactions in the particular context of BECs; the latter
has a number of particularities including e.g., the confinement
that is routinely used to trap and cool the atomic cloud \cite{pethick,book2}.

Multiple dark soliton solutions of the defocusing NLS equation were
first obtained in Ref. \cite{zsd} by means of the inverse scattering
method. Later, an analytical form of a solution of the NLS equation
composed of two dark solitons of different depths and velocities was
found \cite{Blow} (see also the more recent works \cite{AA,gagnon}),
and it was shown that the interaction between dark solitons is
repulsive. Subsequent theoretical studies focused on the
interactions and collisions of dark solitons in the context of
nonlinear optics \cite{Thurston,snyder,Kivshar} and later in BECs
\cite{Huang}, while relevant experimental results (see Refs.
\cite{foursa} for optical dark solitons and \cite{han2,hambcol} for
atomic dark solitons) also examined the interaction between two dark
solitons. However, following the recent experimental methodology of
Ref. \cite{kip}, it is in principle possible to generate multiple
(in fact, in principle, an arbitrary number of) dark solitons: this
can be done by releasing a BEC from a double-well potential into a
harmonic trap within the experimentally accessible, so-called,
dimensionality crossover regime between one-dimension (1D) and
three-dimensions (3D) \cite{kip}. In such a case, it is clear that a
study of multiple matter-wave dark solitons, and their interactions,
should be performed in a theoretical framework that takes into
regard basic features of the pertinent experiment, such as the
effect of dimensionality and the corresponding effective confinement
of the condensate.

In this work, our scope is to analyze this problem, namely the statics and
dynamics of multiple matter-wave dark
solitons in cigar-shaped condensates. Based on the fact that recent atomic
dark soliton experiments were performed
at extremely low temperatures and with sufficiently large number of atoms,
we may safely adopt a mean-field theoretical approach.
In particular, we will perform our analysis in the framework of the
effectively 1D Gross-Pitaevskii (GP) equation with a non-cubic
nonlinearity that was first presented in Ref. \cite{gerbier} and later was also derived and tested in Refs. \cite{Delgado} [this is a distinguishing feature
of our work with respect to
most of the above references which considered dark soliton
interactions in the standard homogeneous defocusing cubic NLS setting].
Our analysis starts by considering the dynamics of multiple dark solitons which is studied as follows.
First, we consider the weakly-interacting limit of the non-cubic GP equation (namely the traditional defocusing NLS model) and,
in the absence of the trap, we derive an effective repulsive potential for the interaction between two solitons.
It is shown that this potential can successfully be used to describe the interactions between ``low-speed'' solitons
(with velocities less than the half of the speed of sound). Such solitons are, ``well-separated'' in the sense
that they always can be identified as distinguishable objects, even at the collision point. Then, using this potential,
we obtain a set of equations of motion for the coordinates of an arbitrary number of solitons. Our approach,
is finally applied to the full problem under consideration (with the
non-cubic nonlinearity and the external harmonic trap), upon
incorporating an effective harmonic potential with a
corresponding characteristic frequency: this is actually the eigenfrequency
of the  first anomalous mode of the system \cite{book2}, corresponding to the
oscillation frequency of a single dark soliton in the trap
(see relevant results in Refs. \cite{Muryshev} and \cite{crossover} for the
purely-1D and dimensionality-crossover regimes).
Such an {\it ad hoc} decomposition of the principal physical mechanisms
affecting the solitons was first introduced in Ref. \cite{kip} (for the case of two symmetrically interacting dark solitons), and
will be validated {\it a posteriori} herein,
by means of direct numerical simulations.

The above  methodology for the study of multiple atomic dark solitons is
directly connected to the
Bogoliubov-de Gennes (BdG) spectrum of excitations of stationary dark soliton
states. The latter are obtained when linearizing around
the nonlinear counterparts of the respective linear states (corresponding to
the eigenmodes of the quantum harmonic oscillator)
\cite{KivsharPLA} and their properties are studied by means of the
well-known BdG equations \cite{book2}. Such an analysis
reveals that the  spectrum of the $n$-th excited state
consists of one zero eigenvalue, $n$ double eigenvalues
(accounted for by the presence of the harmonic trap), and infinitely many
simple ones. In the nonlinear regime, one of
the eigenvalues of each double pair possesses a topological property of,
so-called, {\it negative energy} (in the physical literature) \cite{skryabin}
or {\it negative Krein signature} (in the mathematical literature)
\cite{MacKay};
practically, this means that it becomes structurally unstable, i.e., it becomes complex, upon collision with other eigenvalues.
The eigenvalues with negative Krein signature are actually associated
with the anomalous modes \cite{book2} appearing
in the BdG spectrum. In our case of multiple dark solitons, the number of
anomalous modes in the excitation spectrum
equals to the number of dark solitons \cite{law}, which is in agreement with
the fact that the number of eigenvalues
with negative Krein signature equals to the number of the nodes of the
stationary state \cite{Kapitula}. More generally,
we conjecture (based on the results below for the cases of two- and
three-solitons) that in the case of an $n$-dark
soliton sequence (pertinent to an $n$-th order nonlinear state), the
anomalous modes of the system correspond to the excitation
of the normal modes of the ``dark-soliton lattice''.

The paper is organized as follows. In section II we present the model and make
some general remarks on the theoretical setup.
Section III is devoted to the dynamics of multiple solitons. In particular,
first we analyze the homogeneous weakly-interacting case,
and derive the effective repulsive potential for two solitons undergoing a
symmetric collision. Then, we generalize these results
to include the cases of asymmetric collisions and multiple solitons, as
well as to tackle the full problem, taking into regard the
external harmonic trap and the dimensionality of the condensate. In section
IV we study the stability of the stationary multi-soliton states
via a BdG analysis. We analyze, in particular, the pertinent Bogoliubov
spectra, paying special attention to the anomalous modes of the system.
We illustrate how these anomalous modes correspond to ''normal modes'' of the
``dark-soliton-lattice'', e.g., in-phase and out-of-phase
oscillating dark soliton states. We also predict the onset of dynamical
instabilities due to resonance between the eigenfrequencies of these normal modes and the
excitation frequencies of the background condensate.
We illustrate under what conditions such instabilities
may be observed in future experiments. In Section V we present experimental realizations of
multi-soliton states including a three-soliton state consisting of
two solitons oscillating around a stationary one. Section VI
concludes the paper,
summarizing our findings and presenting some directions of future study.


\section{The model and theoretical setup}

We consider a BEC confined in a highly elongated trap, with longitudinal and transverse confining frequencies
(denoted by $\omega_z$ and $\omega_{\perp}$, respectively) such that $\omega_z \ll \omega_{\perp}$. In this case,
it can be found \cite{gerbier,Delgado} that use of the adiabatic
approximation, in combination
with a variational approach for determining the local transverse chemical
potential, leads to the following effectively 1D GP equation,
\begin {equation}
i\hbar\frac{\partial \psi}{\partial t}=\left[-\frac{\hbar^{2}}{2m}\frac{\partial^{2}}{\partial z^{2}}+ V_{ext}(z)
+\hbar \omega_\perp\sqrt{1+4\alpha |\psi|^2} \right] \psi,
\label{gerbier}
\end {equation}
where $\psi(z,t)$ is the longitudinal part of the condensate's wave function normalized to the number of atoms, i.e.,
$N=\int_{-\infty}^{+\infty} |\psi|^2 dx$, $\alpha$ is the $s$-wave scattering length,
$m$ is the atomic mass, and $V_{ext}(z)$ is the longitudinal part of the external trapping potential, assumed to take the standard
harmonic form $V_{ext}(z) = (1/2)m\omega_z^2 z^2$. As demonstrated in Refs. \cite{Delgado}, Eq. (\ref{gerbier})
provides accurate results in the dimensionality crossover and the Thomas-Fermi limit, thus describing the axial
dynamics of cigar-shaped BECs in a very good approximation to the 3D Gross-Pitaevskii (GP) equation.
Notice that in the weakly-interacting limit, $4a |\psi|^2 \ll 1$, Eq. (\ref{gerbier}) is reduced to the usual
1D GP equation with a cubic nonlinearity, characterized by an effective 1D coupling constant $g_{1D} = 2 a \hbar \omega_{\perp}$.
Equation (\ref{gerbier}) can be expressed in the following dimensionless form,
\begin{equation}
i\frac{\partial \psi}{\partial t}=\left[-\frac{1}{2}\frac{\partial^2}{\partial z^2} + \frac{1}{2}\Omega^2 z^2 +
\sqrt{1+2|\psi|^2}\right]\psi,
\label{1dDelgado}
\end{equation}
where $\Omega \equiv \omega_z/\omega_{\perp}$ is the normalized trap strength. In Eq. (\ref{1dDelgado}),
the density $|\psi|^2$ is measured in units of $2aN$, while length, time and energy are measured, respectively, in units
of the transverse harmonic oscillator length $\alpha_{\perp} \equiv \sqrt{\hbar/m\omega_{\perp}}$, $\omega_{\perp}^{-1}$,
and $\hbar \omega_{\perp}$.

Exact analytical dark soliton solutions of Eq. (\ref{1dDelgado}) are not available. However, following the lines of
Ref. \cite{bass} where a NLS equation with a generalized defocusing nonlinearity was considered, dark soliton
solutions can be found in an implicit form (via a phase-plane analysis) or in an approximate form (via the small-amplitude approximation).
On the other hand, exact analytical dark soliton solutions are available in the above mentioned weakly-interacting limit ($2|\psi|^2 \ll 1$),
and in the absence of the external potential, i.e., for the cubically nonlinear defocusing NLS model.
A single dark soliton solution on top of a background with constant density $n=n_0=\mu$ (with $\mu$ being the chemical potential)
has the form \cite{zsd},
\begin{equation}
\psi(z,t) = \sqrt{n_0}\left[ i\nu +B\tanh(\eta)  \right] \exp(-i\mu t),
\label{single}
\end{equation}
where $\eta = \sqrt{n_0}B(z-\sqrt{n_0}\nu t)$, the parameter $B\equiv \sqrt{1-\nu^2}$ sets the soliton depth, $\sqrt{n_0} B$, while the parameter
$\nu$ sets the soliton velocity, $\sqrt{n_0}\nu$. Note that for $\nu=0$ the dark soliton becomes a stationary kink
(alias ``black'' soliton), while for $\nu=1$ the dark soliton solution (\ref{single}) becomes the background solution. Multiple dark soliton solutions are also available \cite{zsd,Blow,AA}.
In the simplest case of a two-soliton solution, with the two solitons moving with equal velocities, $\nu_1=-\nu_2=\nu$, the wave
function can be expressed as \cite{AA} (see also \cite{gagnon}):
\begin{equation}
\psi(z,t) = \frac{F(z,t)}{G(z,t)}\exp(-i\mu t),
\label{double}
\end{equation}
where $F=2(n_0-2n_{min})\cosh(T)-2n_0 \nu \cosh(Z)+i\sinh(T)$, $G=2\sqrt{n_0}\cosh(T)+2 \sqrt{n_{min}} \cosh(Z)$, while
$Z=2\sqrt{n_0} B z$, $T=2\sqrt{n_{\min}(n_0 - n_{\min})}t$, and $n_{min} =n_0 -n_0 B^2= n_0 \nu^2$ is the minimum density
(i.e., the density at the center of each soliton). This equation will be useful for the analysis of dark soliton
interactions (see next section).

Generally, the single dark soliton, as well as all higher-order dark soliton states, can be obtained in a stationary form
from the {\it non-interacting} (linear) limit of Eq. (\ref{1dDelgado}), corresponding to $N \rightarrow 0$.
In this case, Eq. \ref{1dDelgado} is reduced to a linear Schr{\"o}dinger equation for a confined single-particle state, and
this problem becomes the equation for the quantum harmonic oscillator; the
latter, is characterized by discrete energy levels
and corresponding localized eigenmodes described by the
Hermite-Gauss polynomials \cite{landau}.
Then, in the {\it weakly-interacting} case, where Eq. (\ref{1dDelgado}) becomes the cubic NLS equation,
all these eigenmodes exist for the nonlinear problem as well \cite{KivsharPLA}, describing an analytical continuation
of the linear modes to a set of nonlinear stationary states. Additionally, recent analysis and numerical results \cite{AZ}
suggest that there are no solutions of Eq. (\ref{1dDelgado}) without a linear counterpart. In fact, the effect of interactions
(i.e., the effect of nonlinearity) transforms all higher-order
stationary modes into a sequence of parabolically confined dark solitons
\cite{KivsharPLA}. From the physical point of view, the higher-order
stationary modes exist due to the fact that the repulsion between dark
solitons \cite{Blow,AA,Thurston,snyder,Kivshar} is counter-balanced by the
restoring force induced by the trapping potential.

Below, we are going to analyze the stability of nonlinear modes (namely
stationary multi-dark-soliton states)
of Eq. (\ref{1dDelgado}) by means of the BdG equations. In particular,
first we
identify a numerically exact (up to a prescribed tolerance), stationary soliton state, $\psi_{\rm DS}(x)$, using a fixed point algorithm (e.g., a Newton-Raphson method).
Then, considering small perturbations of this state of the form,
\begin{equation}
\psi(z,t)=\left[\psi_{\rm DS}(z)+ \left(u(z)e^{-i\omega t}+\upsilon^{\ast}(z)e^{i\omega^{\ast} t}\right)\right]e^{-i\mu t},
\label{ansatz}
\end{equation}
(where $\ast$ denotes complex conjugate), we derive from Eq. (\ref{1dDelgado}) the following BdG equations:
\begin{eqnarray}
&&[\hat{H} - \mu + f] u + g\upsilon = \omega u,
\label{BdG1} \\
&&[\hat{H} - \mu + f] \upsilon + gu = -\omega \upsilon,
\label{BdG2}
\end{eqnarray}
where $\hat{H}= -(1/2)\partial_{z}^{2}+(1/2)\Omega^2 z^2$ is the
single particle operator, $\mu$ is the chemical potential and the
functions $f$ and $g$ are given by $f= g +\sqrt{1+4n_{\rm DS}}$, and
$g = 2 \psi_{\rm DS}^2/ \sqrt{1+4n_{\rm DS}}$ (with $n_{\rm DS}
\equiv |\psi_{\rm DS}|^2$). Then, solving Eqs.
(\ref{BdG1})-(\ref{BdG2}), we are going to find the eigenfrequencies
$\omega \equiv \omega_{r}+i \omega_{i}$ and the amplitudes $u$ and
$\upsilon$ of the normal modes of the system. Note that due to the
Hamiltonian nature of the system, if $\omega$ is an eigenfrequency
of the Bogoliubov spectrum, so are $-\omega$, $\omega^{\ast}$ and
$-\omega^{\ast}$. Notice that a linearly stable configuration is
tantamount to $\omega_i =0$, i.e., all eigenfrequencies being real.

The stability of nonlinear modes has already been considered in several works \cite{Carr,Panos,ZA}
in the framework of the 1D GP equation. According to Ref. \cite{Carr} stable nonlinear modes exist, but results
of Ref. \cite{ZA}, obtained near the non-interacting limit, appear to contradict those of Ref. \cite{Carr}.
In fact, in Ref. \cite{ZA} it was claimed that, apart from the first one, all higher-order nonlinear modes
are unstable for repulsive BECs (but can be stabilized by using anharmonic traps). The mechanism of this instability
can be interpreted as follows: in the non-interacting (linear) limit, the
excitation spectrum of the $n$-th mode
possesses $n$ equidistant (due to the harmonic trap) double eigenvalues.
Then, passing into the nonlinear interacting regime,
the negative energy of one member of each of these double eigenfrequency
pairs practically leads to structural instability, i.e., each pair
and its opposite split into a complex quartet of eigenfrequencies.

Below we will show that the above mentioned instability of
higher-order nonlinear modes occurs near the non-interacting limit,
but will cease to exist sufficiently deep inside the nonlinear regime
(i.e., for sufficiently large condensates with large numbers of
atoms $N$). In fact, considering both the second- and the
third-order nonlinear mode of Eq. (\ref{1dDelgado}), we will
demonstrate that they are initially (i.e., for a small number of
atoms, sufficiently close to the linear limit) linearly unstable.
Nevertheless, numerical continuation of the corresponding waveforms
to larger values of $N$, reveals a critical value of the number of
atoms (depending on the anisotropy of the external harmonic trap)
which, if exceeded, all the eigenfrequencies become real and, thus,
the nonlinear stationary state becomes linearly stable.

\section{Dynamics of multiple dark solitons}

\subsection{Multiple soliton interactions -- The homogeneous case}

\begin{figure}
\includegraphics[width=6cm]{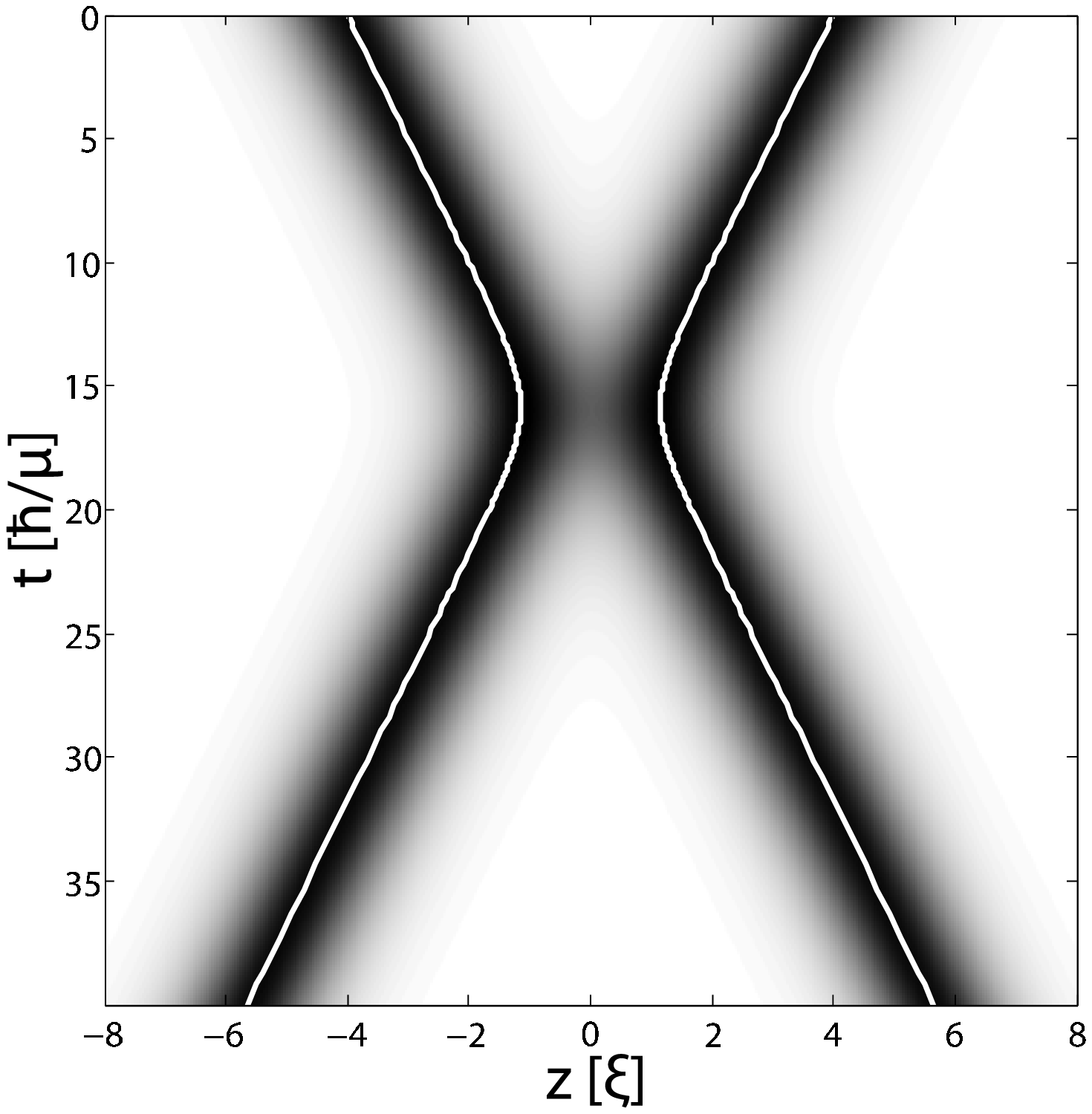}
\includegraphics[width=6cm]{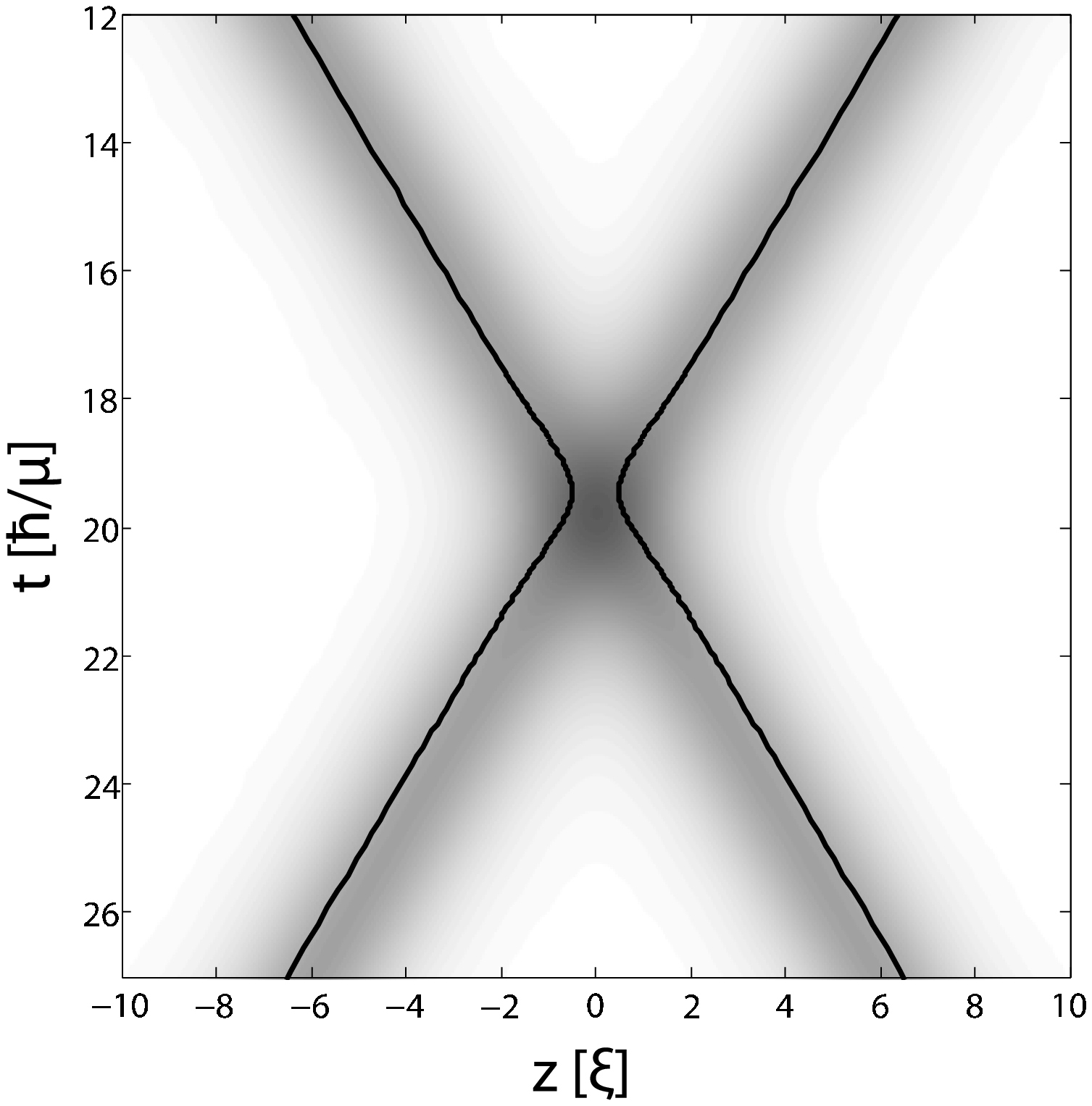}
\includegraphics[width=6cm]{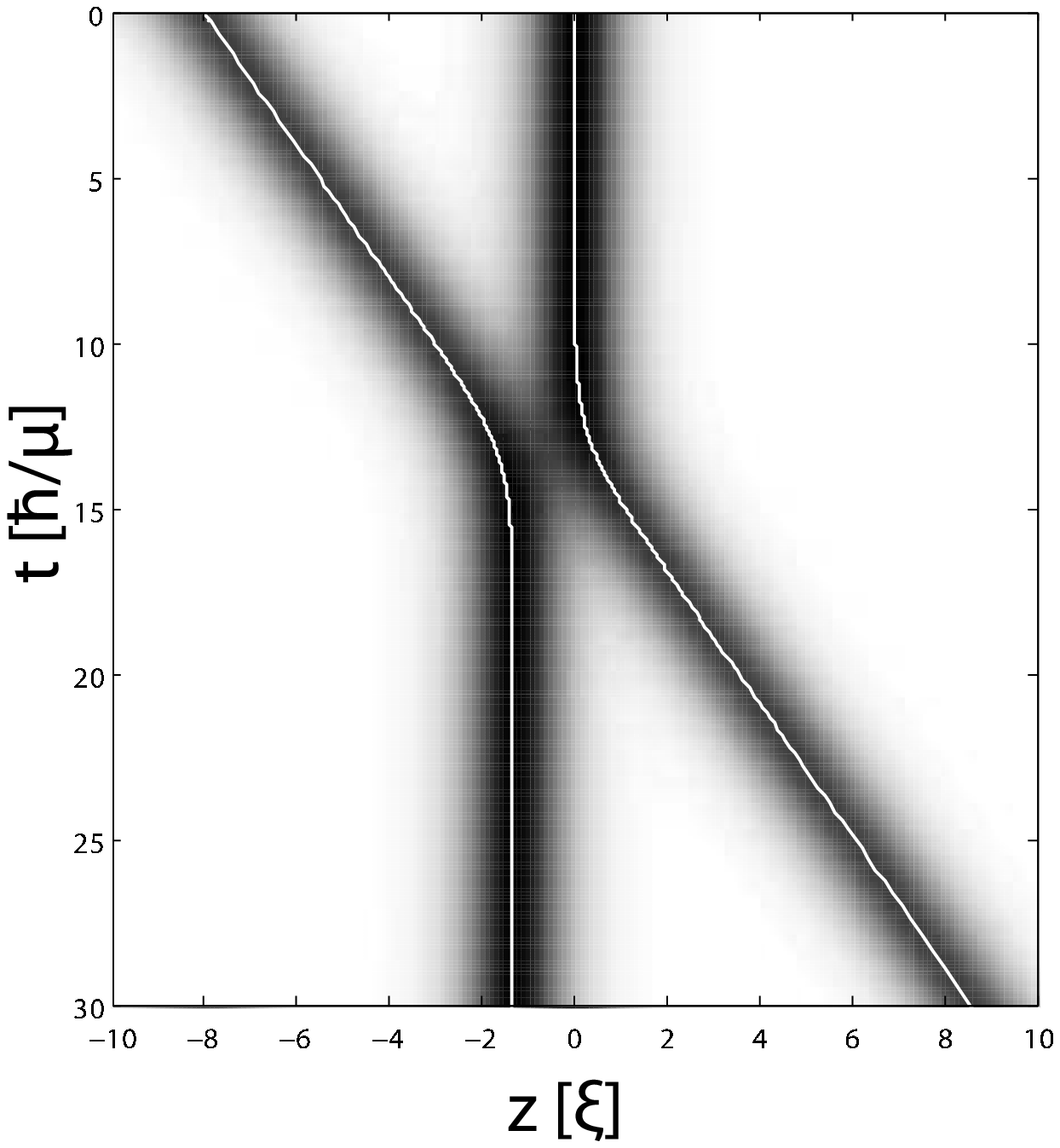}
\includegraphics[width=6cm,height=6.5cm]{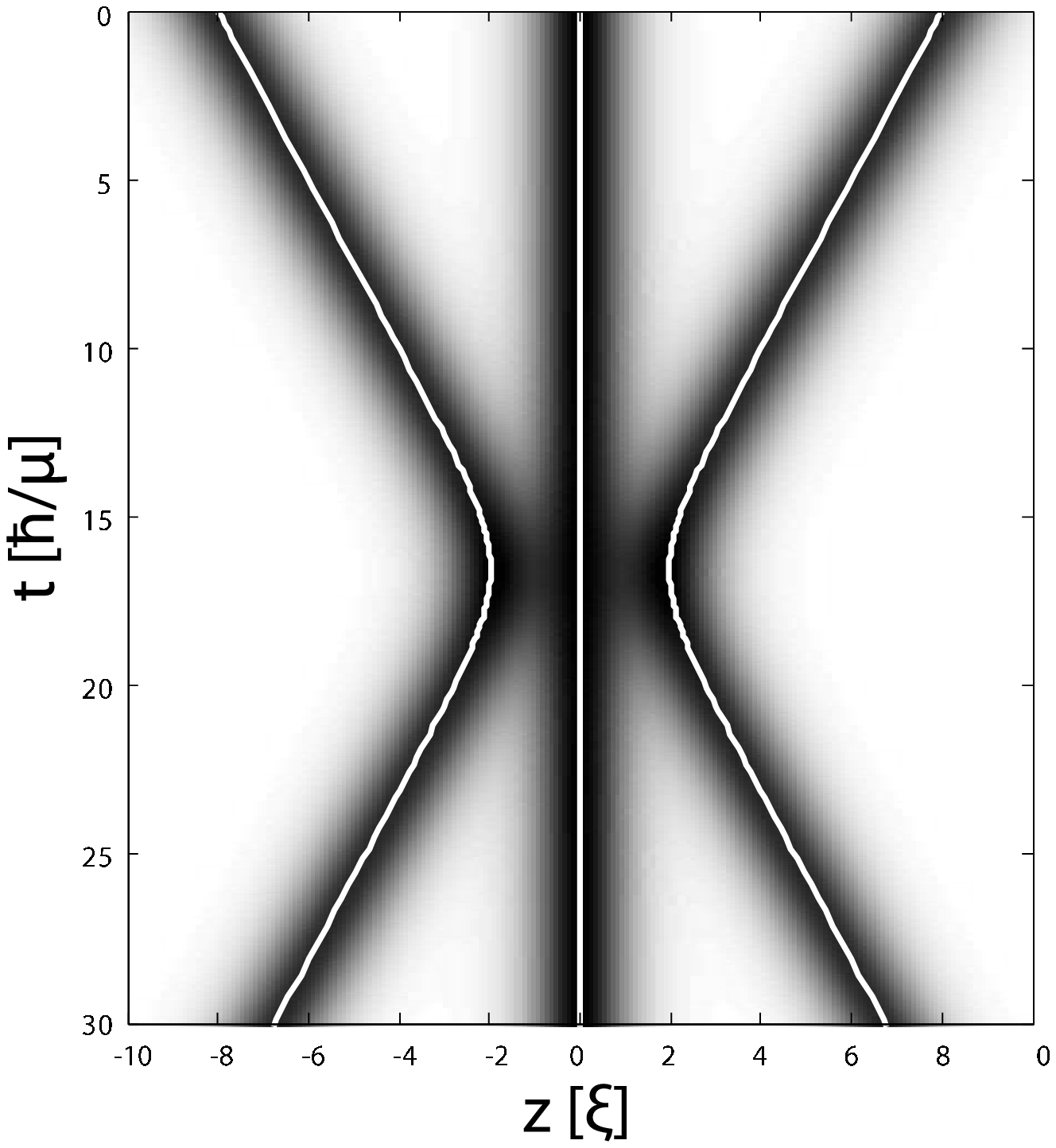}
\caption{Top panels: Soliton trajectories for symmetric two-soliton collisions for different initial velocities:
$\nu=0.2$ (left) and $\nu=0.8$ (right). In the former (latter) case the 
solitons are reflected by (transmitted through) each other.
Bottom panels: Soliton trajectories for an asymmetric two-soliton collision for initial velocities
$\nu_1=0.5$ and $\nu_2=0$ (left), and for a three-soliton collision for initial velocities $\nu_1=-\nu_3 = 0.4$ and $\nu_2=0$ (right).
In all panels, shown are density plots of the wave function carrying the solitons as obtained by
direct numerical integration of the homogeneous NLS equation. The solid lines correspond to the solution of Eq. (\ref{eqm})
(i.e., employing the effective interaction potential). In all cases the normalized chemical potential is $\mu=1$.
}
\label{figcol}
\end{figure}

Let us consider, at first, the simplest possible multi-soliton
state, namely a pair of dark solitons located at $z=\pm z_0$, and
moving with opposite velocities, i.e., $\nu_1 = -\nu_2 = \nu$. Then,
in the framework of the weakly-interacting limit of Eq.
(\ref{1dDelgado}) (and in the absence of an external potential), we
may derive an equation for the trajectory of the soliton coordinate,
$z_{0}$, as a function of time. This can be done upon identifying
the soliton coordinate as the location of the minimum density, and
using the equation $\partial |\psi|^2 / \partial z=0$, with
$\psi(z,t)$ given in Eq. (\ref{double}), to obtain the result:
\begin{equation}
\cosh(2\sqrt{n_0} B z_{0})=\sqrt{\frac{n_0}{n_{min}}}\cosh(T)-2\sqrt{\frac{n_{min}}{n_0}}\frac{1}{\cosh(T)}
\label{traj}
\end{equation}
(recall that $T=2\sqrt{n_{\min}(n_0 -n_{\min})}t$). Then, Eq. (\ref{traj}) can be used for the determination of the distance
$2z^{*}_0$ between the two solitons at the point of their closest proximity (corresponding to $t=0$):
\begin{equation}
z^{*}_{0}=\frac{1}{2\sqrt{n_0-n_{min}}}\cosh^{-1}\left(\sqrt{\frac{n_0}{n_{min}}}-2\sqrt{\frac{n_{min}}{n_0}}\right).
\label{closest}
\end{equation}
This equation shows that the closest proximity distance becomes zero for $n_{min}/n_0 = \nu^2 = 1/4$. Physically, this means that there exists a critical
value of the soliton velocity, namely $\nu_c = 1/2$, which
separates two different regimes: in the first regime,
``high-speed'' solitons with $\nu > \nu_c$ are {\it transmitted}
through each other [their high kinetic energy overcomes the interparticle
repulsion], while in the second regime,
``low-speed'' solitons with $\nu < \nu_c$ are {\it reflected} by each other
[their lower kinetic energy in this regime
is insufficient to overcome the interparticle repulsion].
In fact, as seen in the panels of Fig. \ref{figcol},
the wave function of the low-speed (high-speed) solitons exhibits two separate minima
(a single non-zero minimum) at the collision point, namely $|\psi(z^{*}_{0},t=0)|^2 =0$ ($|\psi(z^{*}_{0},t=0)|^2 \ne 0$).
Note that in the case of the critical velocity $\nu_c=1/2$,
the two-soliton wave function exhibits a single zero minimum at the
collision point.
The above analysis underscores the fact that low-speed solitons are actually ``well-separated'' solitons, in the sense that they can always
be characterized by two individual density minima even at the collision point (the point of their closest proximity). On the contrary, the
high-speed solitons completely overlap at the collision point and, thus,
are not distinguishable during the collision. Thus,
well-separated solitons appear to be reflected by each other and can safely be regarded as hard-sphere-like particles
that interact through an effective repulsive potential (although, as we
will see quantitatively, the description below will be surprisingly
accurate even in the non-well-separated case).

To further elaborate on the above, let us consider the limiting case of extremely slow solitons, namely $n_0/n_{min}=\nu^2 \ll 1/4$,
for which the soliton separation is large for every time, i.e., the closest proximity distance is $z^{*}_{0} \gg 0$.
In this case, the second term in the right-hand side of Eq.(\ref{traj}) is much smaller than the first one for every time
(including $t=0$) and can be ignored. This way, the soliton coordinate can be expressed as:
\begin{equation}
z_{0}=\frac{1}{2\sqrt{n_0} B} \cosh^{-1}\left[\nu^{-1}\cosh(2n_0 \nu B t)\right],
\label{x0}
\end{equation}
which yields the soliton velocities:
\begin{equation}
\frac{dz_{0}}{dt}=\sqrt{n_0}\frac{\sinh(2n_0\nu B t)}{\sqrt{\nu^{-1}\cosh^{2}(2n_0\nu B t)-1}} .
\label{dx00}
\end{equation}
The above equation shows that in the limit $t\rightarrow \pm\infty$, the soliton velocities take the asymptotic values
$dz_0/dt = \pm \sqrt{n_{0}} \nu$, namely the values of the velocities of each individual soliton [see the definition of the
single soliton velocity beneath Eq. (\ref{single})]. On the other hand, at $t=0$, Eq. (\ref{dx00}) yields $dz_0/dt=0$; this
means that as the dark solitons are approaching each other, they become slower, i.e., darker, and at $t=0$
(corresponding to the point of their closest proximity) they become black, remaining at some distance away from each other.
After such a, so-called, head-on ``black collision'' \cite{Huang}, the dark solitons are reflected by each other
and continue their motion in opposite directions.

We now proceed to determine the effective repulsive potential for well-separated dark solitons. This can be done by
determining, at first, an equation of motion for the soliton coordinate: differentiating Eq. (\ref{x0}) twice with respect to time,
and using Eq. (\ref{traj}) (without the second term, which is negligible for well-separated solitons), we obtain
the equation of motion in the form $d^2 z_0/dt^2 =-\partial V(z_0)/\partial z_0$,
with the repulsive potential being given by:
\begin{equation}
V(z_{0})=\frac{1}{2}\frac{n_0 B^2}{\sinh^2(2\sqrt{n_0}Bz_{0})}.
\label{dx01}
\end{equation}
It is worth noting here that since $B=\sqrt{1-\nu^2}$, the above potential
is, in principle, a velocity dependent one.
Note that Eq. (\ref{dx01}) recovers the result obtained in Ref. \cite{Kivshar} by means of a Lagrangian approach
(in that work, the $\sinh$ term in the denominator appears as a $\cosh$ term due to a typographical error \cite{Krol}).

Although the potential of Eq. (\ref{dx01}) is formally applicable only to symmetric collisions,
it can nevertheless be applied also in the case of non-symmetric collisions provided that an ``average depth'' of the two solitons is employed.
In fact, it is possible to generalize this concept for an arbitrary number of solitons, $n$: assuming that the $i$-th soliton
(with $i=1,2,\cdots,n$) is characterized by a darkness $B_i$, velocity $\nu_i= \sqrt{1-B_{i}^2}$, and a position $z_i$,
we may define the average depth
$B_{ij}=(1/2)(B_i + B_j)$ and the
relative coordinate $z_{ij} = (1/2)(z_i - z_j)$ for
solitons $i$ and $j$, and express the interaction potential $V_i$ in the presence of other solitons, as:
\begin{equation}
V_i= \sum_{i\ne j}^{n} \frac{n_0 B_{ij}^2}{2 \sinh^2[\sqrt{n_0} B_{ij}(z_i-z_j)]}.
\label{vj}
\end{equation}
Notice that Eq. (\ref{vj}) is reduced to Eq. (\ref{dx01}) for $\nu_i = -\nu_j= \nu$ (i.e., for $B_i=B_j=B$) and $z_i-z_j=2z_0$.

Using Eq. (\ref{vj}), it is now straightforward to obtain equations of motion for a ``lattice'' consisting of an arbitrary number of dark solitons.
Taking into regard that the Lagrangian $L$ of $n$ interacting solitons is $L=T-V$, where $T\equiv \sum_{i=1}^{n} (1/2) \dot{z_i}^2$
(with $\dot{z_i} \equiv dz_i/dt$) and
$V\equiv \sum_{i=1}^{n} V_i$ are the kinetic and potential energy,
respectively, the Euler-Lagrange equations,
$d(\partial_{\dot{z}_i}L)/dt-\partial_{z_i}L=0$, lead to the following set of
dynamical evolution equations:
\begin{equation}
\ddot{z}_i - \sum_{k=1}^{n}  \left( \frac{\partial^2 V}{\partial z_k \partial \dot{z}_i } \dot{z_k}+
\frac{\partial^2 V}{\partial \dot{z}_k \partial \dot{z}_i} \ddot{z}_k \right) +
\frac{\partial V}{\partial z_i} =0.
\label{eqm}
\end{equation}
These $n$ coupled equations of motion can then be used to calculate the trajectories $z_i(t)$ of $n$-interacting dark solitons. It is worth pointing out
here that in deriving Eqs. (\ref{eqm}), we have attempted to incorporate
the character of the solitary waves as ``deformable particles'' with a
velocity-dependent interaction potential  (cf. Eq. (\ref{dx01}) and
related discussion). This approximation will be tested a posteriori
through the detailed comparison of the particle-based and the GP-based
dynamical results.

\begin{figure}
\includegraphics[width=7cm,height=4.9cm]{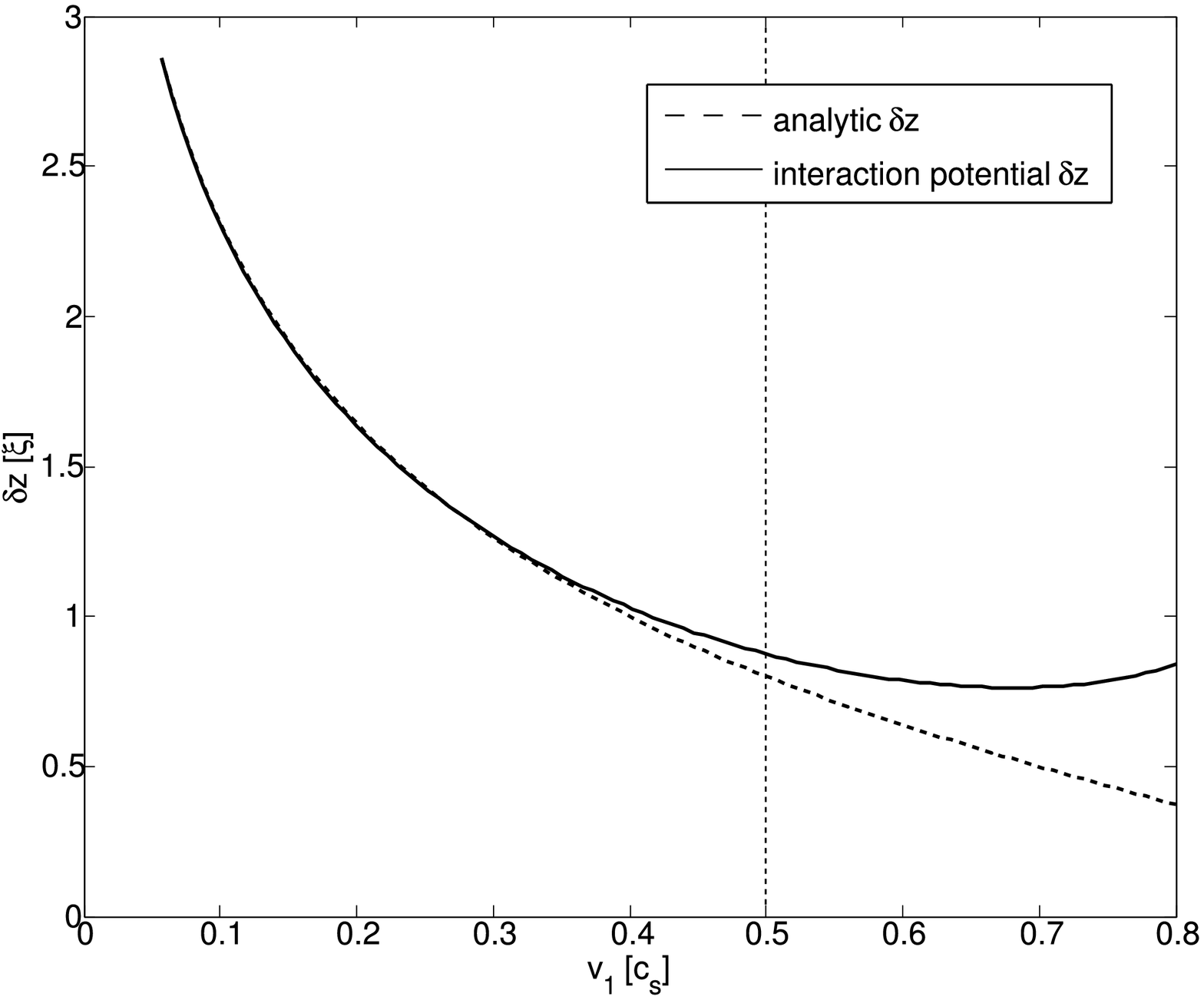}
\includegraphics[width=7cm,height=4.7cm]{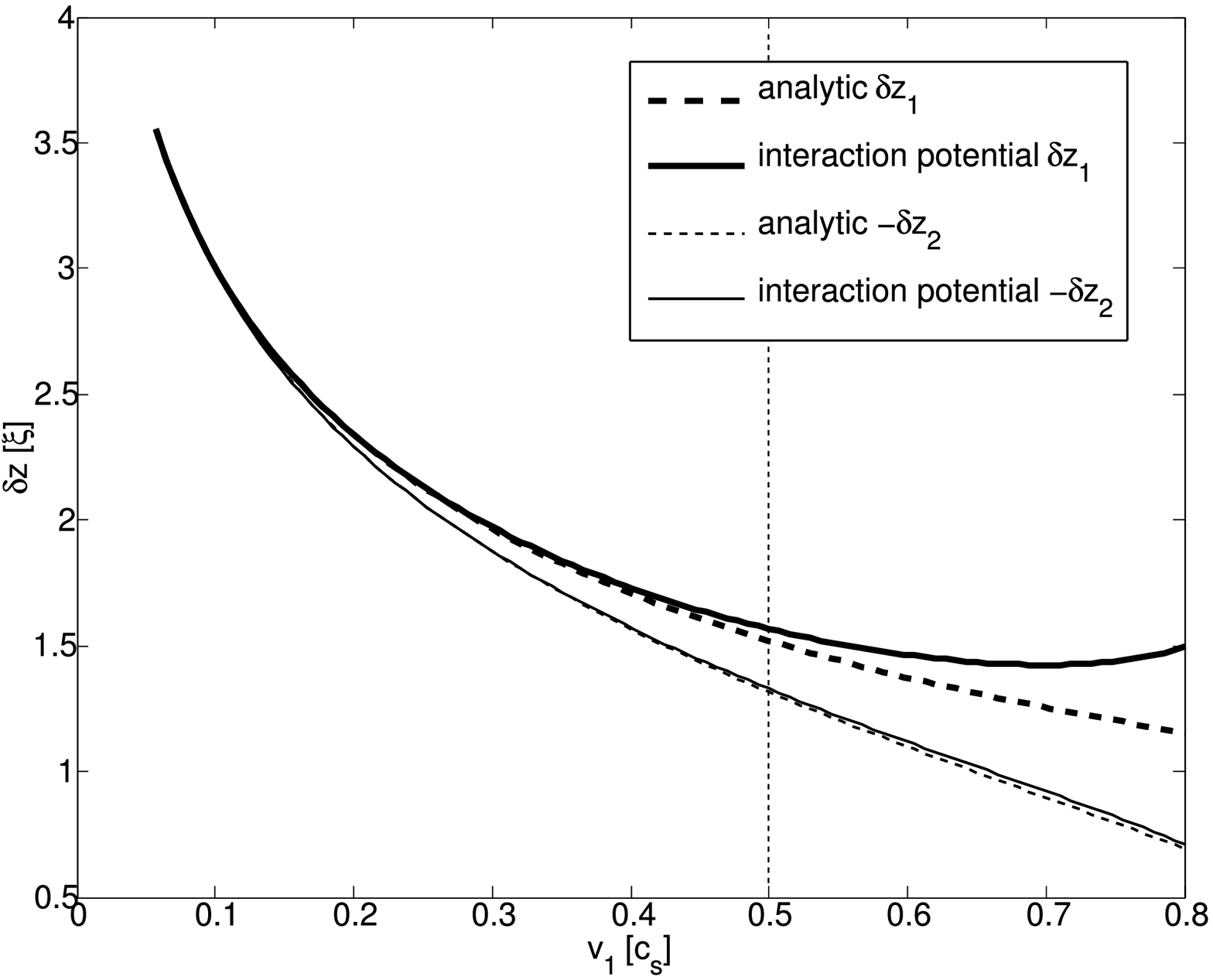}
\caption{Collision-induced phase-shift of the solitons as a function
of the velocity $\nu$. Solid lines depict the analytical result of
Eqs. (\ref{shift1})-(\ref{shift2}) and the dashed lines show the
results obtained from Eq. (\ref{eqm}), which employs the effective
repulsive potential. The agreement between the two approaches is
very good for well-separated solitons, i.e., for $\nu \le 1/2$. The
left panel shows the case of a symmetric collision, and the right
one the case of an asymmetric collision of a moving soliton with
$\nu=\nu_1$ and a stationary soliton (i.e., a black soliton with
$\nu_2 =0$ -- see bottom left panel of Fig. \ref{figcol}). The
dashed vertical line depicts the critical value $\nu=1/2$ and
defines the range of applicability of the approach based on the
effective repulsive potential. } \label{figps}
\end{figure}

We have performed systematic numerical simulations to investigate the range of validity of Eqs. (\ref{eqm}),
both for  cases of symmetric and
asymmetric soliton colisions, as well as both
for  cases of low-speed (well-separated) and high-speed dark solitons.
Various relevant examples are shown in Fig. \ref{figcol}. The simulations
confirm that as long as the dark solitons
are well-separated from each other, i.e., if their depth (velocities) is 
(are) sufficiently large (small), their trajectories
found by means of Eqs. (\ref{eqm}) almost coincide with the ones found by direct numerical integration of the NLS equation.
This excellent agreement can be illustrated not only
qualitatively but also quantitatively: this can be done
upon comparing the exact results for the collision-induced phase-shifts
of the soliton trajectories to the ones
found numerically by means of Eq. (\ref{eqm}). In the case of two solitons, these phase-shifts were calculated analytically
in Ref. \cite{zsd} and have the following form,
\begin{eqnarray}
\delta z_1&=& \frac{1}{2B_1}\ln\frac{(\nu_1-\nu_2)^2 +(B_1+B_2)^2}{(\nu_1-\nu_2)^2 +(B_1-B_2)^2},
\label{shift1} \\
\delta z_2&=& -\frac{1}{2B_2}\ln\frac{(\nu_1-\nu_2)^2 +(B_1+B_2)^2}{(\nu_1-\nu_2)^2 +(B_1-B_2)^2}.
\label{shift2}
\end{eqnarray}
In Fig. \ref{figps} we compare the exact phase-shifts provided by
the above expressions to the ones determined by means of Eq.
(\ref{eqm}), which employ the effective repulsive potential of Eq.
(\ref{dx01}). Both cases of a symmetric (top panels) and an
asymmetric (bottom left panel) collision are shown; it is clearly
observed that the agreement between the two approaches is very good
for soliton velocities $\nu \le 1/2$, i.e., for well-separated dark
solitons. Notice that in the case of an asymmetric collision, $\nu_1
\ne \nu_2$, the two shifts are not equal, $|\delta z_1| \ne |\delta
z_2|$, while in the case of a symmetric collision, $\nu_1 =
-\nu_2=\nu$ (and, thus, $B_1=B_2=B$), the phase shifts become equal,
$|\delta z_1| = |\delta z_2| = (2B)^{-1}\ln(1+B^2/\nu^2)$. 

\subsection{Dynamics and interactions of multiple solitons in the trap}

The above analysis of soliton interactions in the homogeneous case is of use
in the inhomogeneous case as well.
In particular, here we will consider multiple dark solitons in the presence of an harmonic trap, also taking into regard that
the condensate is cigar-shaped, so that the proper model is Eq. (\ref{1dDelgado}) (rather than its weakly-interacting limiting case,
i.e., the usual cubic NLS equation considered above). In such an experimentally relevant situation, we may employ the theoretical approach
adopted in Ref. \cite{kip} and use an
interaction potential for dark solitons of the form:
\begin{equation}
V_i^{eff} = V_{ext}^{eff}(z_i) + V_i(z_i, \dot{z}_i),
\label{actpot}
\end{equation}
where $V_i(z_i, \dot{z}_i)$ is the interaction potential of Eq. (\ref{vj}) and the effective trapping potential $V_{ext}^{eff}$ is given by:
\begin{equation}
V_{ext}^{eff}(z_i) = \frac{1}{2}\omega_{osc}^2 z_i^2.
\label{veff}
\end{equation}
The underlying assumption within this decomposition is that dark soliton
is an effective particle moving under the combined influence of external
forces from the confining potential and from the other solitons
within the configuration. Each of these individual forces has an associated
potential and hence the total force and associated motion stem from
the combination of these potentials. A more subtle assumption is
that while the effect of dimensionality on a single soliton is
captured in the effective oscillation frequency $\omega_{osc}$ discussed
in detail in the next paragraph, the tail-tail interaction of the waves is well
approximated by its NLS counterpart. These assumptions will be
validated a posteriori through our comparisons between
theoretical and numerical results below.

Here, $\omega_{osc}$ is the oscillation frequency of a single dark soliton in the harmonic trap, which coincides with the lowest anomalous
mode of the system \cite{crossover}. In the purely 1D regime,
and for sufficiently large number of atoms (i.e., in the so-called
Thomas-Fermi regime \cite{pethick,book2})
the soliton oscillation frequency is $\omega_{osc} = \Omega/\sqrt{2}$,
where $\Omega$ is the normalized trap strength.
This can be derived either by analyzing the dynamics of the
dark soliton in the trap (see, e.g., \cite{revnonlin} and references therein)
or by means of a BdG analysis \cite{Muryshev}. In the case of cigar-shaped BECs under consideration, the oscillation frequency is upshifted
\cite{crossover} and, generally, takes values in the interval $\Omega/\sqrt{2} < \omega_{osc} < \Omega$.

Based on the above discussion, the interaction potential of Eq. (\ref{veff})
takes into account both the effective harmonic trap (including the
dimensionality
of the system), $V_{ext}^{eff}(z_i)$, and the inter-soliton interaction
potential, $V_i(z_i, \dot{z}_i)$ (derived for the homogeneous 1D regime).
This potential has already been successfully used in Ref. \cite{kip}, where
the experimental findings for the symmetric collisions between two dark
solitons were found to be in excellent agreement with the corresponding
theoretical results. Here, we will show that the approach based on the use
of the effective potential of Eq. (\ref{actpot}) can also be generalized to
the case of asymmetric collisions. Such a case can also be investigated
experimentally in the context of the experimental setup of
\cite{kip}. In particular, in \cite{kip}
an even number of dark solitons was created by merging two condensates
initially prepared in a double-well trap, with a zero-phase
difference between them. In principle, it is also easy to create an
odd number of solitons, upon introducing a non-zero phase difference
between the wells, which would lead to an asymmetric evolution pattern
of the solitary waves. If, furthermore, the phase difference is exactly
equal to $\pi$, then a stationary (black) dark soliton is created
exactly at the center of the harmonic trap. In fact,
this procedure has already been used in relevant interference
experiments (not directly connected, however, to dark solitons) \cite{jo}.

\begin{figure}
\includegraphics[width=7cm,height=5.2cm]{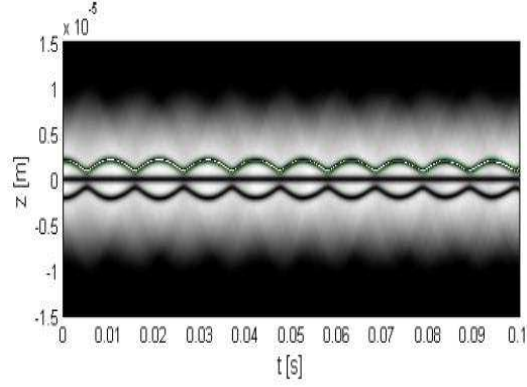}
\caption{(Color online) A configuration of three dark solitons, with the central one being stationary and the other two
oscillating with a frequency $\omega_{osc}=0.904 \Omega$. Shown are a contour plot depicting the evolution of the density according to
Eq. (\ref{1dDelgado}) illustrating the path of the density minimum of one of
the oscillating solitons, as well as
the respective trajectory calculated by the equation of motion employing the effective potential of Eq. (\ref{actpot}) (solid line).
The parameter values are $N = 2000$, $\omega_z= 2\pi \times 53$ Hz, $\omega_{\perp}= 2\pi \times = 890$ Hz (i.e., $\Omega \approx 0.06$)
and initial displacement from the trap center $\delta z_0 = 2\mu$m.
}
\label{tria}
\end{figure}

In Fig. \ref{tria} we show an example of an asymmetric collision between a stationary dark soliton (with $\nu_2=0$) at a trap center
and a pair of oscillating solitons (with $\nu_1=-\nu_3$) in a cigar-shaped condensate confined in a trap of strength $\Omega \approx 0.06$.
The figure shows the evolution of the density as obtained by direct numerical
integration of Eq. (\ref{1dDelgado}).
Additionally, the figure shows the trajectory of one of the solitary
waves as computed via the equation of motion that makes use of the
effective potential of Eq. (\ref{actpot}) (solid green line)
[the other moving dark soliton is the mirror image of the one shown around
$z=0$, while the third is, by symmetry, constrained to stay precisely
at $z=0$]. It is clear
that the agreement between  Eq. (\ref{1dDelgado}) and the reduced
particle picture  is excellent.

We complete the analysis of this section by studying the case of dark solitons performing small oscillations around their equilibrium positions.
In fact, we are going to use the Lagrangian approach devised above to connect the oscillation frequency obtained by the multi-soliton dynamics to
the eigenfrequencies of the anomalous modes of the stationary soliton states that will be obtained by a BdG analysis in the next section.
In that respect, it is relevant to consider the simplest case of two well-separated solitons, which are assumed to be almost black
(i.e., $B_1=B_2 \approx 1$). In such a case, the Lagrangian takes the form,
\begin{equation}
L=\frac{1}{2}\dot{z_1}^2+\frac{1}{2}\dot{z_2}^2-\frac{1}{2}\omega_{osc}^{2}z_{1}^2-\frac{1}{2}\omega_{osc}^{2}z_{2}^2
-\frac{n_0}{\sinh^2(\sqrt{n_{0}}(z_2-z_1))}.
\label{Lag2ds}
\end{equation}
Then, using the Euler-Lagrange equations, and replacing the hyperbolic function $\sinh$ by its exponential asymptote in the case
under consideration, i.e., for $|z_2-z_1|>>0$,
the following equations of motion are obtained
\begin{eqnarray}
\ddot{z}_1 & = &-8n_{0}^{3/2}e^{-2n_{0}^{1/2} (z_2-z_1)}-\omega^2_{osc}z_1 \\ \nonumber
\ddot{z}_2 & = & 8n_{0}^{3/2}e^{-2n_{0}^{1/2} (z_2-z_1)}-\omega^2_{osc}z_2.
\label{system1}
\end{eqnarray}
The fixed points, $Z_1$, $Z_2$ of the above system can be easily found by setting the left-hand side equal to zero. The result is:
\begin{equation}
Z=Z_2=-Z_1=\frac{1}{4\sqrt{n_{0}}}w \left(\frac{32n_{0}^2}{\omega_{osc}^2}\right),
\label{equil2ds}
\end{equation}
where $w(\eta)$ is the Lambert's w function defined as the inverse of
$\eta(w)= w e^{w}$.
Then, considering small deviations ($\eta_1, \eta_2$) from the equilibrium positions $Z_1$, $Z_2$, we can Taylor expand the
interaction potential keeping only the lowest-order term and, this way, derive the following linearized equations of motion:
\begin{eqnarray}
\ddot{\eta}_1 & = & 16n_{0}^{2}e^{-4\sqrt{n_{0}}Z}(\eta_2-\eta_1)-\omega^2_{osc}\eta_1, \\ \nonumber
\ddot{\eta}_2 & = -&16n_{0}^{2}e^{-4\sqrt{n_{0}}Z}(\eta_2-\eta_1)-\omega^2_{osc}\eta_2.
\label{system1linear}
\end{eqnarray}
Let us now consider the normal modes of the system and seek solutions of the form $\eta_i=\eta_{i0}e^{i\omega t}, i=1,2$,
where $\omega$ is the common oscillation frequency of both dark solitons. Then, substituting this ansatz into
Eqs. (\ref{system1linear}), we rewrite the equations of motion as matrix eigenvalue equation, namely:
\begin{equation*}
-\omega^2\mathbf\eta = \left(
\begin{array}{cc}
-\omega_{osc}^2-16n_{0}^2e^{-4\sqrt{n_{0}}Z} & 16n_{0}^2e^{-4\sqrt{n_{0}}Z}  \\
16n_{0}^2e^{-4\sqrt{n_{0}}Z} & -\omega_{osc}^2-16n_{0}^2e^{-4\sqrt{n_{0}}Z}  \\
\end{array} \right)\mathbf{\eta}.
\end{equation*}
To this end, it is possible to obtain from the above system the
characteristic frequency $\omega_{1}=\omega_{osc}$,
which corresponds to in-phase oscillations of the two dark solitons,
as well as the frequency $\omega_{2}$, which
corresponds to out-of-phase oscillations of the two solitons.
The latter is given by:
\begin{equation}
\omega_{2}=\sqrt{\omega_{osc}^2+32n_{0}^2e^{-4\sqrt{n_{0}}Z}}.
\label{outofphase}
\end{equation}

The above procedure can also be applied to the case of three, almost black, solitons ($B_i \approx 1, i=1,2,3$) considering
only nearest-neighbour interactions. In this case, the equilibrium positions are given by the following expressions
\begin{equation}
Z_2=0, \tilde{Z}=Z_3=-Z_1=\frac{1}{2\sqrt{n_{0}}} w\left(\frac{16n_{0}^2}{\omega_{osc}^2}\right),
\label{equil3ds}
\end{equation}
while the three characteristic frequencies which correspond to the three normal modes of the system are the following,
\begin{eqnarray}
\omega_{1}&=&\omega_{osc}, \\ \nonumber
\omega_{2}&=&\sqrt{\omega_{osc}^2+16n_{0}^2e^{-2\sqrt{n_{0}}\tilde{Z}}} \\ \nonumber
\omega_{3}&=&\sqrt{\omega_{osc}^2+48n_{0}^2e^{-2\sqrt{n_{0}}\tilde{Z}}}.
\label{3ds_normal_modes}
\end{eqnarray}
In the following section, we will elaborate on the investigation of the stability of the multiple soliton states and the derivation
of their anomalous modes.

\section{Stability of stationary multi-soliton states}

Having examined the location of the solitary waves and their
oscillation eigenmodes and eigenfrequencies in an analytical
form, we now turn to the numerical investigation of such
stationary multi-soliton states, and to their corresponding
BdG spectrum. We carry out the relevant computations first
for the two-dark-soliton state and subsequently for the
three-dark-soliton state.

\subsection{The two-dark-soliton state}

We start by considering the simplest possible stationary multi-soliton state, namely the second-order nonlinear mode of Eq. (\ref{1dDelgado}).
In the linear limit of $N \rightarrow 0$, this state corresponds to the second-excited state of the quantum harmonic oscillator.
The excitation spectrum of this state contains a zero eigenvalue (corresponding to the Goldstone mode), two double eigenfrequencies located at $\Omega$ and $2\Omega$, as well as infinitely many simple eigenfrequencies. Below we will analyze the excitation spectrum
in the nonlinear regime (i.e., when the number of atoms $N$ is increased)
focusing, in particular, on the two double eigenvalue pairs mentioned above.
Notice that in our numerical results (see below) we fix $\omega_{\perp}$ to the typical value $\omega_{\perp}=2\pi \times 400$ Hz.

\begin{figure}
\includegraphics[width=8cm]{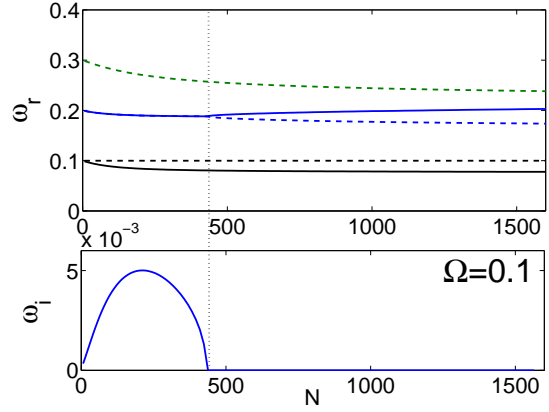}
\caption{(Color online)
Top panel: the real part $\omega_r$ of the lowest eigenfrequencies of the second nonlinear mode as a function of the number
of atoms $N$ (for a harmonic trap with $\Omega=0.1$).
Eigenfrequencies possessing negative Krein signature, namely eigenfrequencies of the two anomalous modes, are depicted
by solid lines. Eigenfrequencies possessing positive Krein signature are depicted by dashed lines, with the lowest ones
starting from $\omega_r = 0.1$ and $\omega_r = 0.2$ at $N=0$ corresponding, respectively, to the dipole and quadrupole modes.
Bottom panel: the imaginary part $\omega_i$ of the eigenfrequency pair starting from $\omega_r= 0.2$ at $N=0$ as a function of
the number of atoms $N$. The vertical dotted line at the critical value $N=438$ indicates the splitting of the complex pair into real eigenfrequencies.
}
\label{fig1}
\end{figure}

First, we consider the two eigenfrequencies located at $\Omega$ (for
$N=0$) which, in the nonlinear regime, obtain opposite Krein
signature. In particular, one of them has positive Krein signature
and corresponds to the {\it dipole mode} \cite{pethick,book2}, while
the second one has negative Krein signature, i.e., the integral of
the norm $\times$ energy product, $\int (|u|^2 -|v|^2)\omega  dx $
(in our units), is negative. In other words, in the nonlinear regime
this eigenvalue becomes the eigenfrequency $\omega_1$ of one of the
two anomalous modes of the system (recall that the number of
anomalous modes in the excitation spectrum is the same as the number
of dark solitons or nodes in the relevant waveform). Notice that
both eigenfrequencies originating from $\Omega$ in the linear limit
[see dashed (solid) lines for the one with positive (negative) Krein
signature in the top panel of Fig. \ref{fig1}] are real for every
value of the number of atoms $N$, indicating the absence of any
instability. In fact, the dipolar mode, per the relevant symmetry
\cite{pethick,book2}, remains fixed at $\omega=\Omega$, while the
anomalous mode's frequency in line with the dependence of the single
dark soliton mode anomalous mode frequency \cite{crossover}.

Next, we consider the eigenvalue pair located at $2\Omega$ (for
$N=0$). Similarly to the previous case, these eigenfrequencies
obtain opposite Krein signature: one eigenvalue has positive Krein
signature and corresponds to the background condensate's {\it
quadrupole mode} \cite{pethick,book2}, while the second one has
negative Krein signature, thus being the eigenfrequency $\omega_2$
of the second anomalous mode of the system. An important difference
from the previous case is that this second pair of double
eigenfrequencies does become complex --  see bottom panel of Fig.
\ref{fig1}, where the imaginary part of these eigenfrequencies is
shown as a function of $N$. This implies that the respective
nonlinear stationary state is unstable for sufficiently small atom
numbers. Nevertheless, the instability occurs only near the linear
limit (as was also predicted in Ref. \cite{ZA}). However, as seen in
the bottom panel of Fig \ref{fig1}, when the number of atoms exceeds
a critical value, namely $N=438$ for a trap strength $\Omega=0.1$,
all the eigenfrequencies become real and the nonlinear state becomes
linearly stable.

We have generally found that the larger the number of atoms, and the
stronger the anisotropy of the harmonic trap,
the more stable the configuration with the two stationary dark
solitons is. For example, for $\Omega=0.35$
this state is unstable up to the number of atoms $N=1067$, while
for $\Omega\approx0.1$ the instability occurs for
very small condensates, with number of atoms $N<500$. Notice that a similar
behavior was also found in the framework
of the 1D GP equation considered in Ref. \cite{ZA} (results not shown here).
Additionally, regarding the connection of
our analysis with experimental observations, we note that
within the parameter range of relevance to the recent experiment of
\cite{kip}, we found the two-dark soliton state to be linearly stable.

Let us now return to the excitation spectrum of Fig. \ref{fig1} and focus on the eigenfrequencies possessing
negative Krein signature (see solid lines in the top panel of
Fig. \ref{fig1}), namely the two anomalous modes.
In the case of two dark solitons under consideration, the physical
significance of these two anomalous modes
has been discussed in Ref. \cite{law}. More specifically, excitation of the
anomalous mode with the smallest eigenfrequency, $\omega_1$,
gives rise to an {\it in-phase} oscillation, i.e., the two dark solitons
move towards the same direction without
changing their relative spatial separation. On the other hand,
excitation of the anomalous mode with the largest eigenfrequency, $\omega_2$,
gives rise to an {\it out-of-phase} oscillation, i.e., the two dark solitons
move in opposite directions with the same
velocity and undergo a head-on collision. It should be pointed out
that in the context of the analysis of the previous section,
this information is also encompassed within the eigenvectors
$\eta_{i0}$ of the small perturbations around the stationary
multi-soliton state. In particular, the two possibilities correspond,
respectively, to $\eta_{20}=\eta_{10}$ and $\eta_{20}=-\eta_{10}$.

\begin{figure}
\includegraphics[width=8cm]{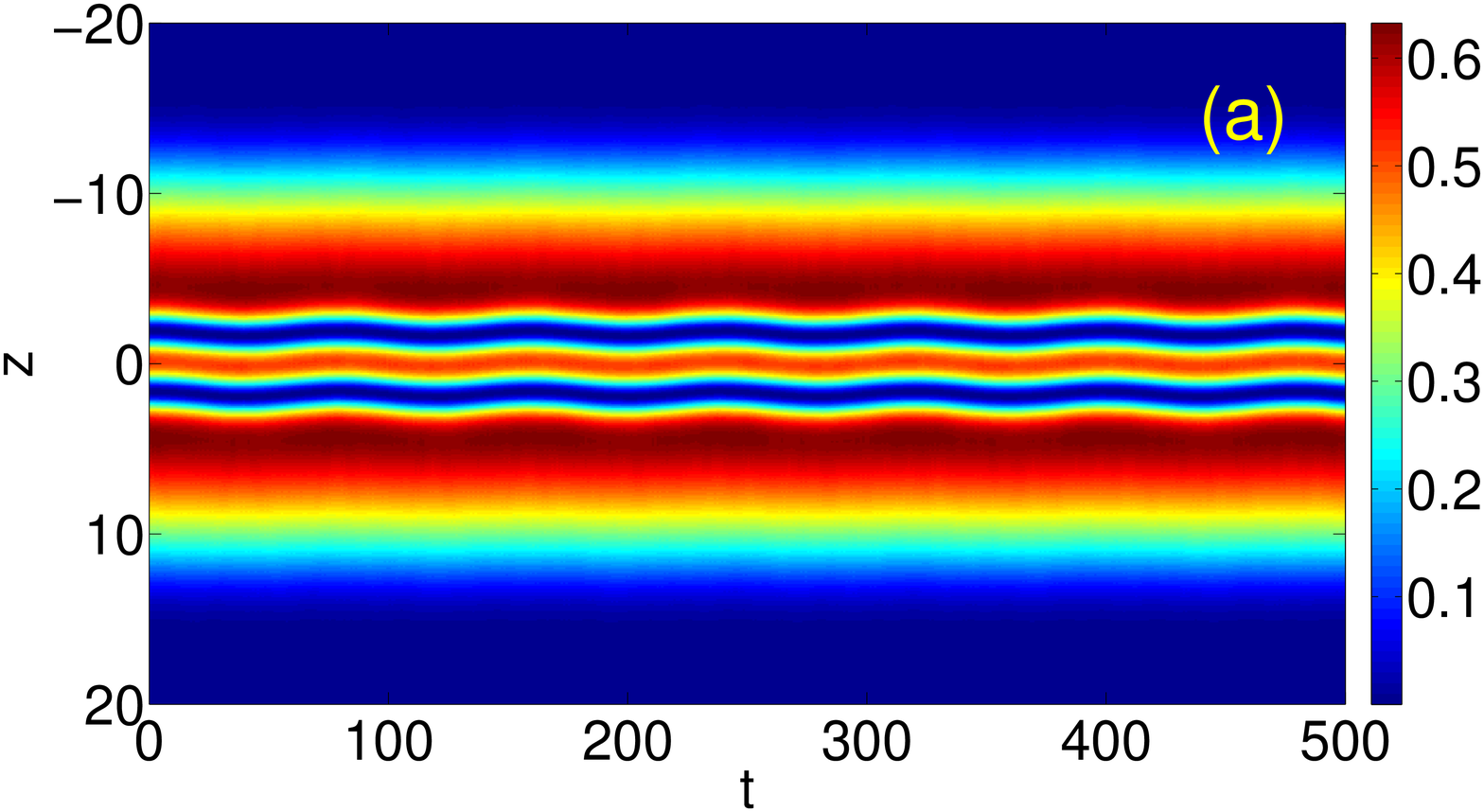}
\includegraphics[width=8cm]{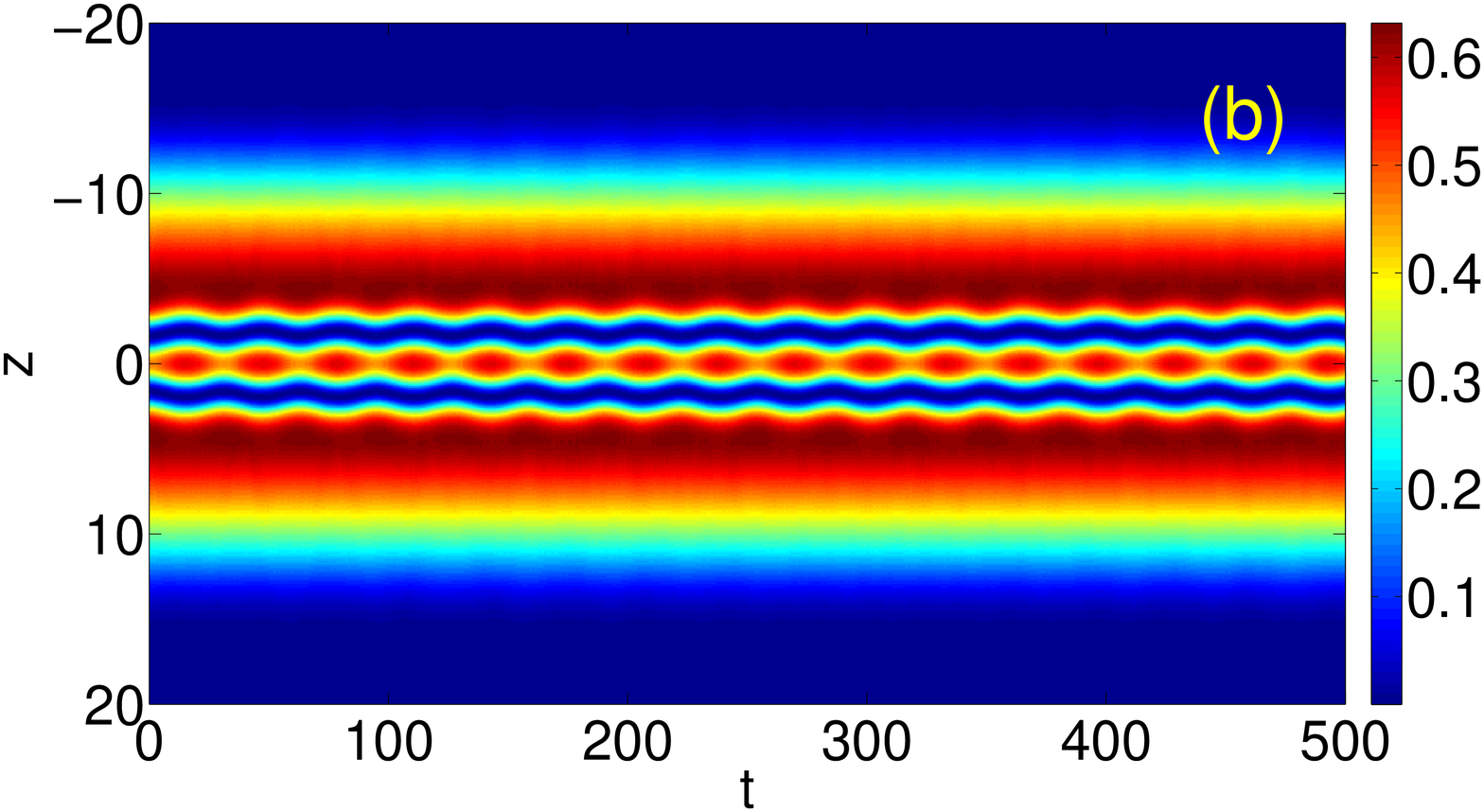}
\includegraphics[width=8cm]{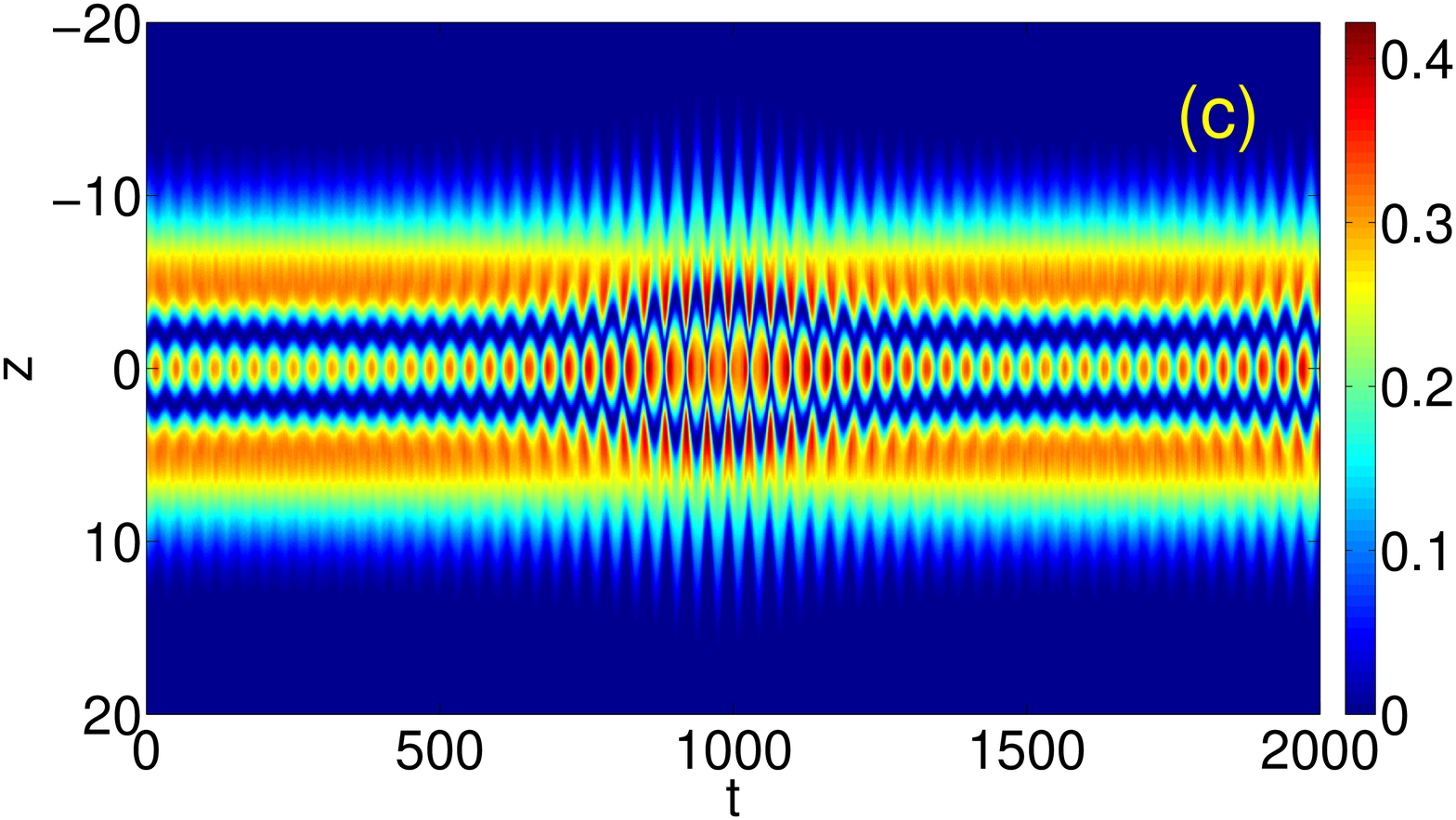}
\caption{(Color online)
Spatio-temporal evolution of the condensate density after the excitation of the two anomalous modes. Panels (a),(b) show the
in-phase and out-of-phase oscillatory motion of the two dark solitons (for $\Omega=0.1$ and $N=1000$); the oscillation
frequencies are identical to the eigenfrequencies of the first and second
anomalous modes, respectively.
Panel (c) shows the manifestation of the dynamical instability
(for $\Omega=0.1$ and $N\approx 400$)
due to the collision of the second anomalous mode with the quadrupole mode:
the motion of the solitons excites
the quadrupole mode of the system, resulting in a breathing behavior of the
condensate.
}
\label{fig2e}
\end{figure}

The correspondence of the anomalous modes with the normal modes of
the two-dark soliton state is confirmed
by direct numerical simulations. In particular, we have numerically
integrated Eq. (\ref{1dDelgado}) with initial condition
the nonlinear stationary mode excited by the corresponding anomalous modes, i.e., $\psi(x;t=0)=\psi_{\rm DS}(x)+ u(x)+\upsilon^{\ast}(x)$.
The results (for parameter values
$\Omega =0.1$ and $N\approx 1000$) are shown in
panels (a), (b) of Fig. \ref{fig2e}.
In Fig. \ref{fig2e}(a)  (Fig. \ref{fig2e}(b)), the excitation of the first (second) anomalous mode results in an in-phase (out-of-phase) oscillatory
motion of the two dark solitons, with the characteristic eigenfrequency of the first (second) anomalous mode, i.e., $\omega_1 = 0.0784$ ($\omega_2 = 0.199$).
Calculating numerically the maximum density of the stationary state and the equilibrium positions of the dark solitons, we find that $n_0=0.623$ and $Z=1.78\pm0.1$. This allows us to make a comparison with the ones calculated
analytically based on the theoretical approach of the previous section.
Using Eq. (\ref{equil2ds}), with $\omega_{osc}=\omega_1$, and Eq. (\ref{outofphase}) we find that $Z=1.85$ and
$\omega_{2}=0.205$, which differ only $3\%$ from the
corresponding numerically obtained values.

Finally, it is worth investigating the manifestation of the instability predicted above for small condensates,
due the ``collision'' of the second anomalous mode with the
quadrupole mode. In Fig. \ref{fig2e}(c), we show
the evolution of the density of a condensate with $N \approx 400$ confined in a trap with strength $\Omega =0.1$ (as before).
For these parameter values, the eigenfrequencies of the two anomalous modes are $\omega_1= 0.081$ and $\omega_2=0.188+0.0025i$.
The numerical integration of Eq. (\ref{1dDelgado}) reveals that although the initial evolution of the density roughly follows the
one observed in Fig. \ref{fig2e}(b) (up to $t \approx 700$),
the instability eventually manifests itself:
the soliton motion excites the quadrupole mode of the system, and this
excitation results in a breathing behavior of the BEC
(see the bottom panel of Fig. \ref{fig2e}).

\subsection{The three-dark-soliton state}

\begin{figure}
\includegraphics[width=8cm]{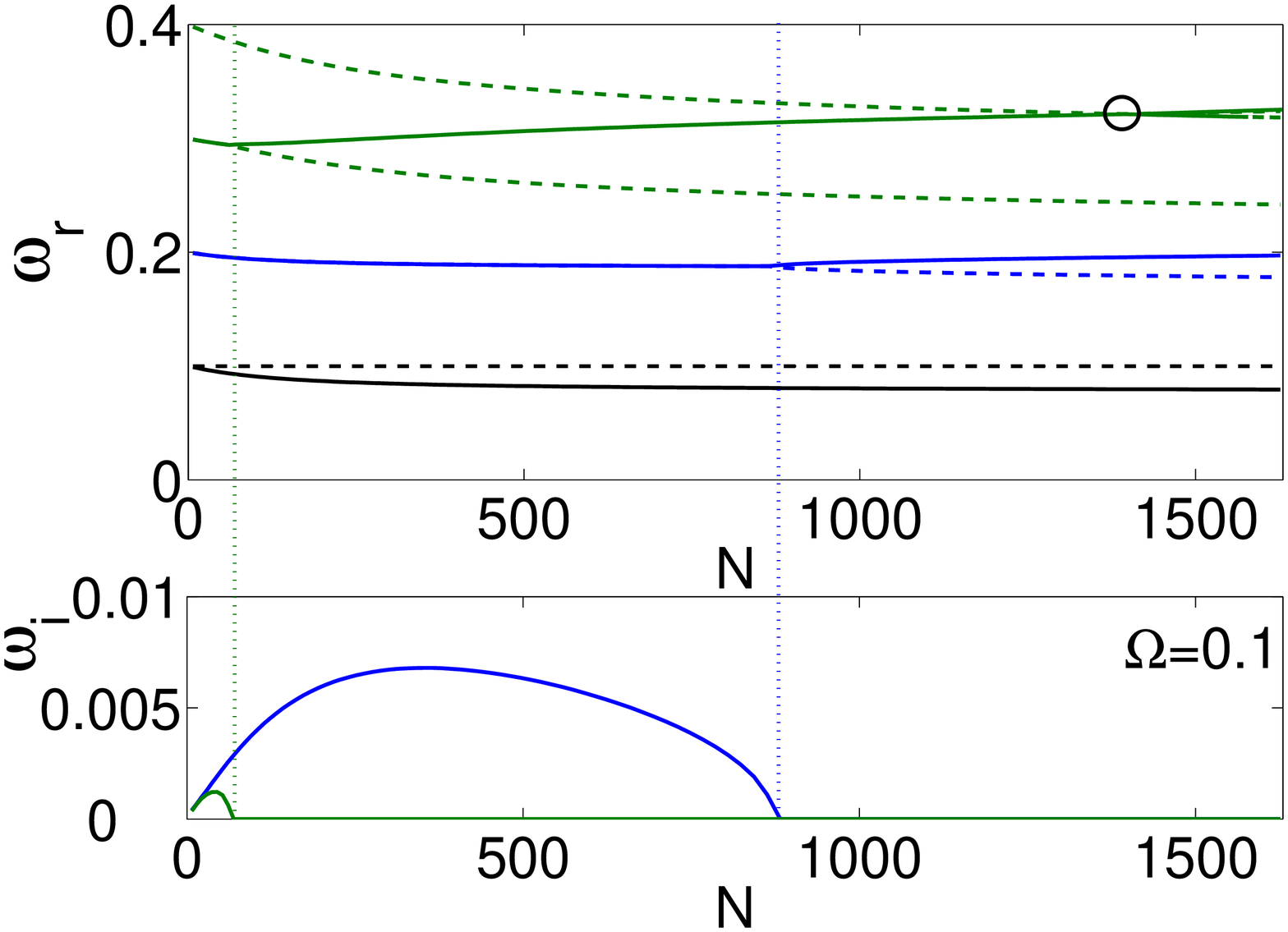}
\caption{(Color online)
Top panel: The real part of the lowest eigenfrequencies of the third
excited mode (i.e., three-dark-soliton state) as a function of the
number of atoms $N$
(for a harmonic trap with $\Omega=0.1$). The notation follows the one
used in Fig. \ref{fig1}.
The bold circle in the top panel indicates the collision of
eigenfrequencies
of opposite Krein signatures which does not lead to instability (see
details in the text).
Bottom panel: The imaginary parts of the the eigenfrequency pairs starting
from $\omega_r= 0.2$ (blue) and $\omega_r= 0.3$ (green) at $N=0$ as a
function of the number of atoms $N$.
The vertical dotted lines indicate the splitting of the respective complex
pairs into
real ones. After the second splitting, all the eigenfrequencies become real
and the three-soliton state is stabilized.
}
\label{fig3ds1}
\end{figure}

Let us now consider the third-order nonlinear mode of Eq. (\ref{1dDelgado}).
In the linear limit, the excitation
spectrum of this state consists of a zero eigenvalue, three double
eigenfrequencies located at $\Omega$, $2\Omega$, and $3\Omega$,
as well as infinitely many simple ones. In the nonlinear regime, and
similarly to the previous case, each of the
the three aforementioned pairs obtain opposite Krein signature.
In Fig. \ref{fig3ds1} we show the real (top panel)
and imaginary (bottom panel) parts of the lowest eigenfrequencies as a
function of the number of atoms $N$ (for a trap strength
$\Omega=0.1$ as before). In the top panel, the eigenfrequencies depicted by
dashed (solid) lines correspond to
ones with with positive (negative) Krein signature. Regarding the
ones with positive Krein signature, we note that the lowest
ones, starting from $\omega_r=0.1$ and $0.2$ for $N=0$, correspond to the dipole and quadrupole modes, respectively. As before, the system's ability to
sustain dipolar oscillations for all $N$ with a frequency $\Omega$
preserves the dipolar frequency at $\omega_r=0.1$ throughout the
relevant figure and precludes the possibility of a quartet-inducing
collision with the lowest in-phase-oscillation anomalous mode of the
system.

We now focus on the two upper double pairs
(and their anomalous modes), located at $2\Omega$ and $3\Omega$ in the linear
limit. These become complex in
the nonlinear regime. As is observed in Fig. \ref{fig3ds1}, the upper pair (starting from $3\Omega$ for $N=0$)
splits into real eigenfrequencies at very small values of the number of
atoms $N$, while the lower
pair (starting from $2\Omega$ at $N=0$) remains complex for larger values of
$N$. For the assumed trap strength $\Omega=0.1$,
the complex eigenfrequencies become real beyond the critical value
of $N\approx 880$ and, thus, the nonlinear mode becomes
dynamically stable. Here, it is worth mentioning that this state remains
stable for the values of  $N \gtrsim 880$ considered herein,
although at $N\approx 1395$ another collision appears: indeed,
at a point marked by a circle in the top panel of Fig. \ref{fig3ds1},
the eigenfrequencies starting from $\omega_r=0.3$ and $0.4$ for $N=0$, which possess opposite Krein signature, cross each other.
Nevertheless, this collision does not lead to instability because the
eigenmodes associated with these eigenfrequencies remain
orthogonal at the collision point. This happens due to the opposite parity
of the colliding eigenmodes \cite{ZA}.

\begin{figure}[tbp]
\includegraphics[width=8cm]{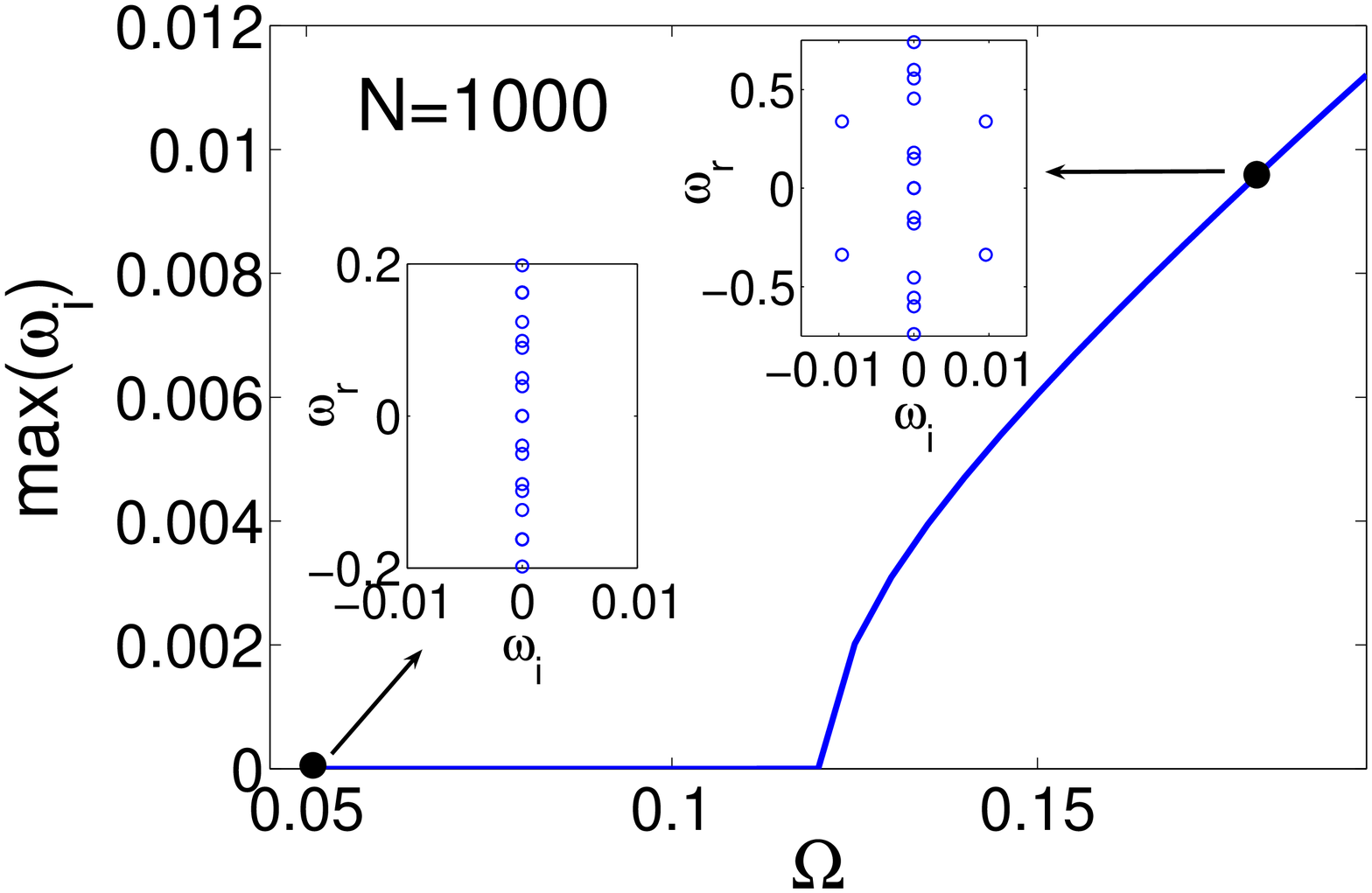}
\caption{(Color online)
The maximum of the imaginary part of the eigenfrequencies of the third
excited state for a fixed number of atoms ($N=1000$) as a function
of the trap strength $\Omega$. The insets show the spectral plane for a stable ($\Omega=0.05$)
and an unstable ($\Omega=0.18$) case. In the latter case, the complex
eigenfrequencies originate
from the collision of the second anomalous mode with the quadrupole mode.
}
\label{fig3ds2}
\end{figure}

So far, to investigate the stability of the three-dark soliton state
(as well as the two-soliton state considered in the previous section)
we kept the trap strength $\Omega$ fixed and varied the number of atoms $N$.
It is also worthwhile (and experimentally relevant) to reverse the
procedure, i.e.,
to keep the number of atoms fixed, at $N=1000$, and vary the harmonic trap
strength $\Omega$, as shown in
Fig. \ref{fig3ds2}. In this figure, it is readily observed that the nonlinear
state remains stable up to the critical value
$\Omega=0.12$. Beyond this value, the eigenfrequency of the second anomalous mode collides with that of the quadrupole mode and becomes
complex. Typical examples of the spectral plane, for both stable and unstable
cases, are shown in the insets of Fig. \ref{fig3ds2}.
For the stable case of $\Omega=0.05$, the eigenfrequencies of the
anomalous modes are found to be
$\omega_{1}=0.039$, $\omega_{2}=0.0988$, and $\omega_{3}=0.1626$, while for the unstable case of $\Omega=0.18$
the respective values are $\omega_{1}=0.1489$ and $\omega_{2}=0.5994$; notice that there exist also two complex eigenfrequencies
at $\omega_{2}=0.3379 \pm 0.01 i$, stemming from the above mentioned collision.
Once again, we calculate numerically the maximum density of the state,
$n_0=0.3817$, and  the equilibrium positions of the dark
solitons: $Z_2=0, Z_3=-Z_1=4.3\pm0.1$.
Using Eq. (\ref{equil3ds}), with $\omega_{osc}=\omega_1$, we find that 
$Z=4.54$, in good agreement with the numerically obtained, while the 
remaining two normal mode frequencies given by 
Eq. (\ref{3ds_normal_modes}) are found to be
$\omega_{2}=0.1$, $\omega_{3}=0.1647$ and differ by 
less than $2\%$ from the ones obtained by the BdG analysis.

\begin{figure}
\includegraphics[width=8cm]{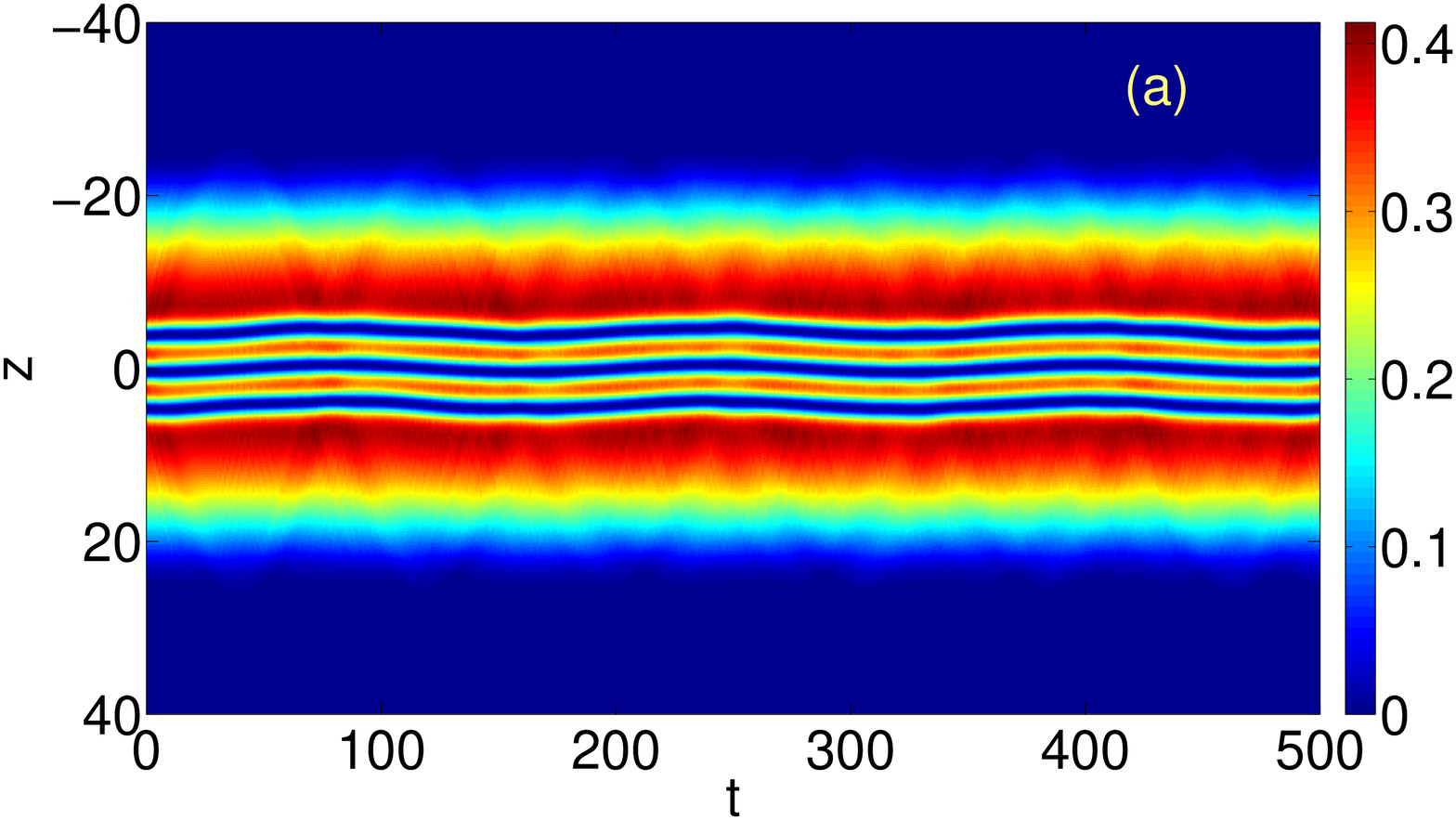}
\includegraphics[width=8cm]{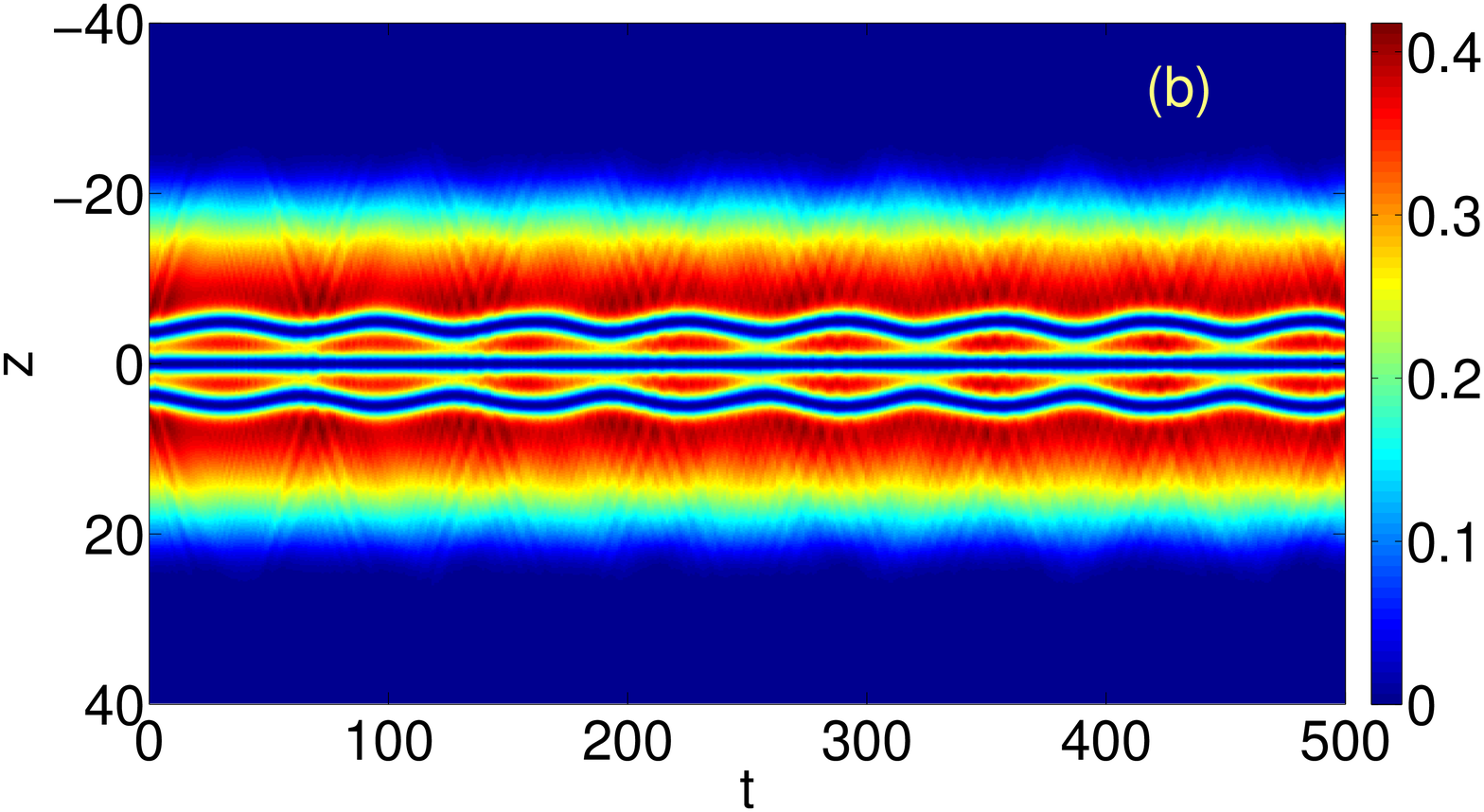}
\includegraphics[width=8cm]{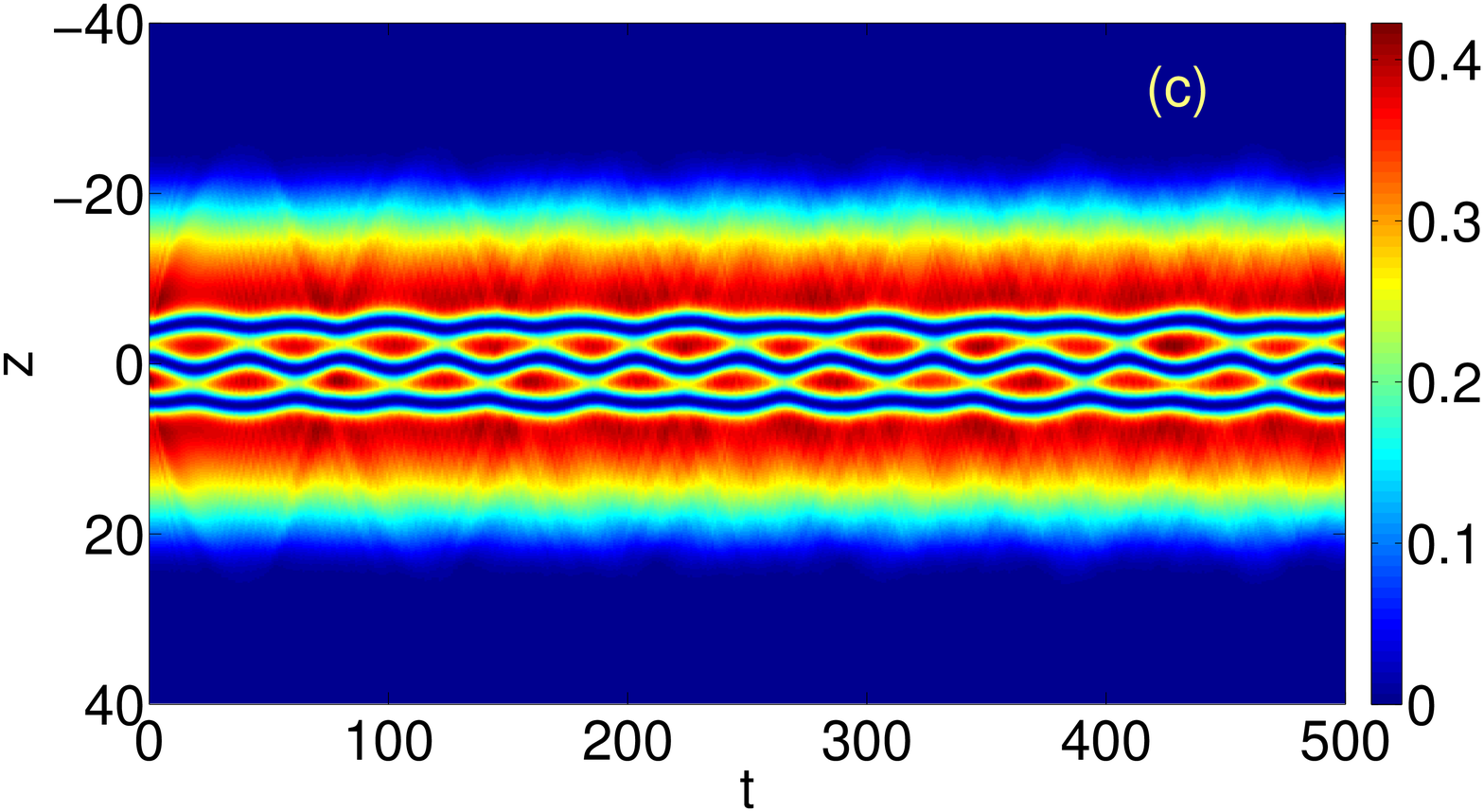}
\includegraphics[width=8cm]{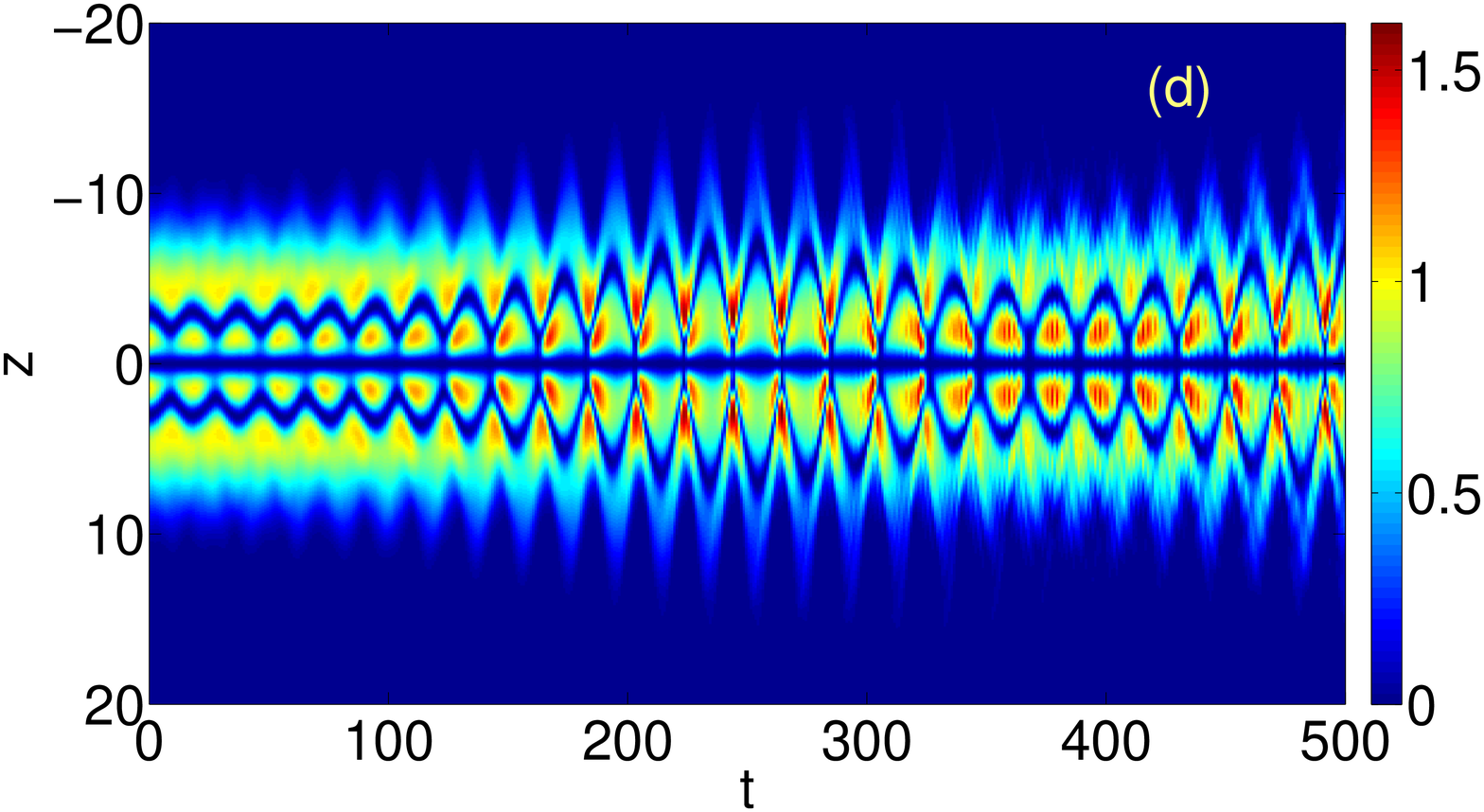}
\caption{(Color online)
Spatio-temporal evolution of the condensate density after the excitation of
the three anomalous modes.
In panels (a)-(c) stable configurations with $\Omega=0.05$, $N=1000$ are
shown: in (a) excitation of the first anomalous mode
results in a configuration where the three dark solitons move together
in-phase; in (b) excitation of the second anomalous mode
results in a configuration where the outer solitons move in
opposite directions (each being the mirror image of the other
around $z=0$) while the middle one is at rest;
in (c) excitation of the third anomalous mode results in a configuration
where the center soliton is moving in
opposite direction to the outer dark solitons. Panel (d) shows the
excitation of the mode associated with a complex eigenfrequency
for the unstable case with $\Omega=0.18$
and $N=1000$. Here, a dynamical instability manifests itself due to the
collision of the second anomalous mode with the quadrupole mode:
the motion of the dark solitons excites the quadrupole mode of the system,
resulting in a breathing behaviour of the entire condensate.
}
\label{fig3ds3}
\end{figure}

Similarly to the two-soliton state, we now present results of direct numerical integration of Eq. (\ref{1dDelgado}) with an
initial condition given by
the third nonlinear mode excited by the corresponding anomalous modes.
First we study the stable case,
with $\Omega= 0.05$ and $N=1000$, and then the unstable case,
with $\Omega=0.18$ and $N=1000$.
In the stable case, when the first anomalous mode is excited, the three dark solitons perform an in-phase oscillation
with the characteristic eigenfrequency of the corresponding anomalous
mode -- see Fig. \ref{fig3ds3}(a).
The excitation of the second anomalous mode results in the following
configuration: the two outer solitons
are moving in opposite directions (symmetrically around $z=0$),
while the center soliton is a stationary one -- see Fig. \ref{fig3ds3}(b).
As mentioned in the previous section, this configuration may be
experimentally observed following the experimental procedure
of \cite{kip}, i.e., by merging two condensates initially prepared in
a double-well trap,
with a $\pi$-phase difference between them. On the other hand,
when the third anomalous mode is excited,
the two outer solitons are oscillating in-phase while the center soliton
is oscillating out-of-phase with respect to the outer ones --
see Fig. \ref{fig3ds3}(c)).
Finally, in the unstable case, we use as an initial condition the third
nonlinear state excited by the mode associated with
the complex eigenfrequencies. As seen in Fig. \ref{fig3ds3}(d),
initially the outer dark solitons are moving out-of-phase, following
the configuration observed in Fig. \ref{fig3ds3}(b). Nevertheless,
similarly to the two-dark soliton state,
the motion of the solitons gradually excites the quadrupole mode of the
system, resulting in a breathing behaviour of the condensate,
signalling the manifestation of the relevant dynamical instability.
Clearly, in such a case, the oscillation amplitude of the dark
solitons is not constant anymore.

\section{Experimental creation of multiple atomic dark solitons}

Dark solitons can be created experimentally,
e.g., by the method of matter-wave interference,
which can be considered as a form of density engineering. This method makes use of the fact
that an interference pattern
in the presence of interatomic interactions
generates a train of dark solitons \cite{Scott}. Our experimental
realization of density engineering involves two BECs, initially
prepared in a double-well potential. This double well potential is
created by the superposition of a crossed optical dipole trap and a
one-dimensional optical lattice \cite{Alb}.
Removing the optical lattice leads to the merger of the two initially separated BECs in
the harmonic trap and to the subsequent creation of an interference
pattern which generates the solitons. The experimental procedure is
described in \cite{kip}. Two oscillating and colliding solitons
were observed in that experiment.
In the following we will extend this scheme to the preparation of a single and of multiple
dark solitons in a harmonic trap.\\

\subsection{Controlling the created macroscopically excited state}

\begin{figure}[tbp]
\includegraphics[width=12cm]{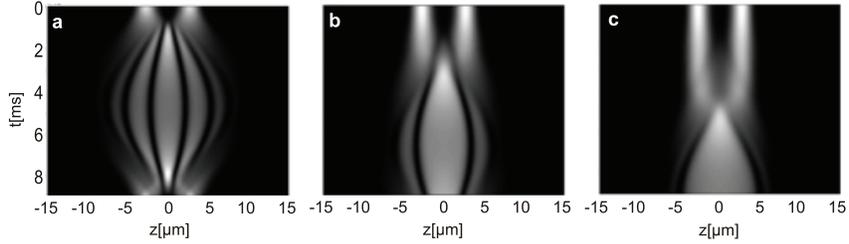}
\caption{Increasing the distance between the created solitons can be
realized by increasing the ramping-down time of the optical
lattice $\tau_{OL}$, as shown by numerical simulations of Eq. (\ref{1dDelgado}): a) $\tau_{OL}=0ms$, b)
$\tau_{OL}=2ms$, c) $\tau_{OL}=4ms$} \label{SW_ramp}
\end{figure}

The fringe spacing of a matter-wave interference pattern
depends on the momentum of the two merging atom clouds. Therefore, it is
possible to vary the number of created solitons in this process by controlling the relative velocity.
An increase of the relative velocity between the atom clouds lowers
the distance between the created solitons and leads to the creation
of additional solitons further away from the trap center. This can
be realized by
changing the ramping-down time of the optical lattice $\tau_{OL}$, as shown in
Fig. \ref{SW_ramp}. The number and distance of created solitons can also be controlled by the aspect ratio of the trap $\omega_z/\omega_{\perp}$ or by lowering the number of atoms.
As thermal fluctuations of the initial phase directly translate into
position fluctuations of the dark soliton train, care has to be taken
to keep this phase as rigid as possible. Since the phase
fluctuations scale as $\delta\Phi \propto k_B T/E_j$ (with $E_j$
being the tunneling coupling energy of a double-well system), a
careful control of the potential height, i.e. $E_j$, allows to
stabilize the phase and thus leads to negligible position
fluctuations. This also limits the achievable soliton distances. In
our experiment we can realize $\delta\Phi \approx 0.06/(2\pi)$. 

\begin{figure}[h]
    \centering
    \includegraphics[width=0.6\textwidth,keepaspectratio]{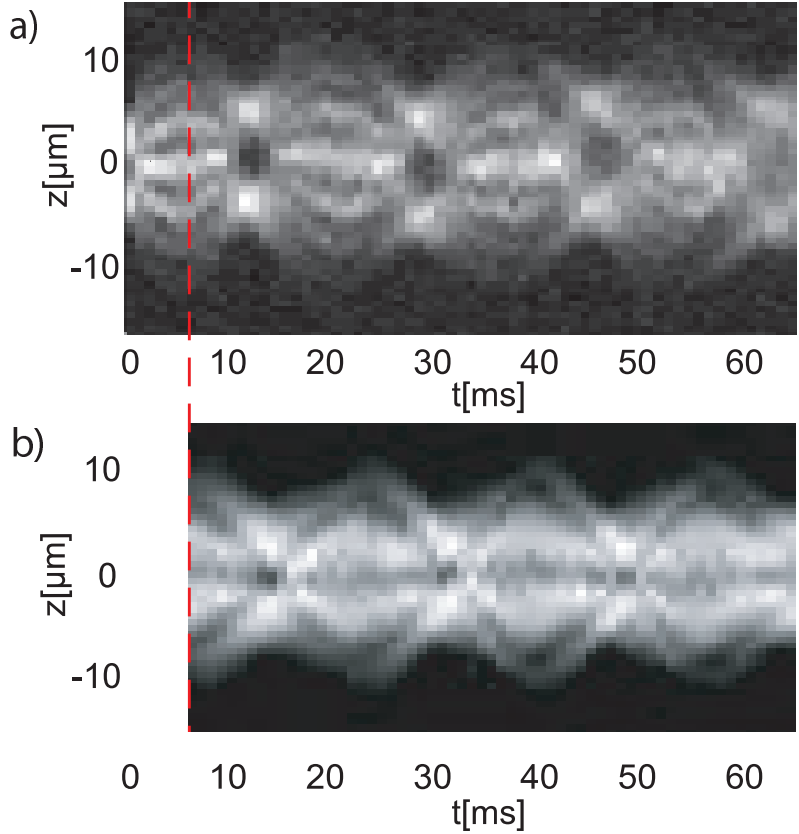}
    \caption{a) Observation of the oscillation of four dark solitons including the
    creation process. The evolution is averaged over 10 experimental runs. The point in time where
    the creation process of the solitons is finished is marked by the dashed line.
    b) Experimental observation of three dark solitons in a harmonic trap averaged over 16 runs.
    The creation process of the solitons is not shown in this case.
    The soliton in the center of the trap is at rest (black soliton)
    whereas the two outer ones oscillate.
    The time evolution plots were obtained by
    integrating the images over their transverse axis, meaning that each vertical line
    shows the longitudinal density of the BEC at a certain point in time.
    \label{3DS_4DS}
     }
\end{figure}

The successful experimental realization of four solitons plus two additional weak
ones with extremely high oscillation amplitude is shown in Fig.
\ref{3DS_4DS}(a). The illustrated oscillation dynamics, recorded with
a time resolution of $1$ms, includes the creation process of the
solitons starting at the initial double well potential and including
the ramping down of the optical lattice. The creation process ends at the
dashed line, where the final value of the longitudinal trapping
frequency is reached after a suitable ramping down from the value necessary to
obtain the double well potential. 

A matter-wave interference pattern depends on the relative phase
difference $\Delta\phi$ of the two merging atom clouds.
If $\Delta\phi=0$ a symmetric pattern with an even number of
solitons is produced. A small phase difference leads to an
asymmetric evolution pattern of the created solitons, while a phase
difference close to $\pi$ leads to the creation of an additional
soliton between the other ones
meaning that an odd number of solitons is produced.
Such phase difference can be created by changing the symmetry of the
potential, which results in an energy difference between the levels
of the two wells of the double-well potential.
Maintaining this asymmetry for a certain hold time
accumulates a phase difference between the two BECs.
In a simple approximation, the phase difference is given by $\Delta
\phi\approx\Delta E/\hbar \cdot t$. By adapting the asymmetry and
the time of phase accumulation before releasing the two condensates
from the double well, arbitrary phase differences can be achieved.
Especially interesting is the case were the initial phase difference
is exactly $\pi$. Then,
a black (stationary) soliton
is created at the center of the trap, between the oscillating
ones.
Shifting the symmetry of the potential experimentally is realized by
shifting the second beam of the dipole trap with respect to the
optical lattice.

Fig. \ref{3DS_4DS}(b) shows the experimental realization of three dark
solitons in a harmonic trap created by the above discussed method.
The trap frequencies used in the experiment are
$\nu_z=36.1\pm0.25{\rm Hz}$ (longitudinal frequency),
$\nu_{\perp}=407.5\pm40.8{\rm Hz}$ (transverse frequency). The mean
number of atoms in the BEC is $N=1570\pm146$. In this measurement
the height of the optical lattice is ramped down linearly on a timescale of
$\tau_{OL}=2{\rm ms}$. The final value of the longitudinal trapping
frequency $\nu_z$ is reached after ramping down within $7$ms from the
value of $\nu_z^{initial}=63{\rm Hz}$ necessary for obtaining the double
well potential \cite{nu_z}.
In the performed experiment the oscillation amplitude of the two
outer solitons, $A_{osc}=(21\pm 0.6)\xi$, is
relatively large.
Therefore,
the oscillation frequency is only moderately increased by the
soliton-soliton interaction in this case:
$\nu_d/\nu_z=0.775\pm0.006$. A numerical simulation yields
$\nu_d/\nu_z=0.761$ in good agreement with the experimental result.

\begin{figure}[h]
    \centering
    \includegraphics[width=0.8\textwidth,keepaspectratio]{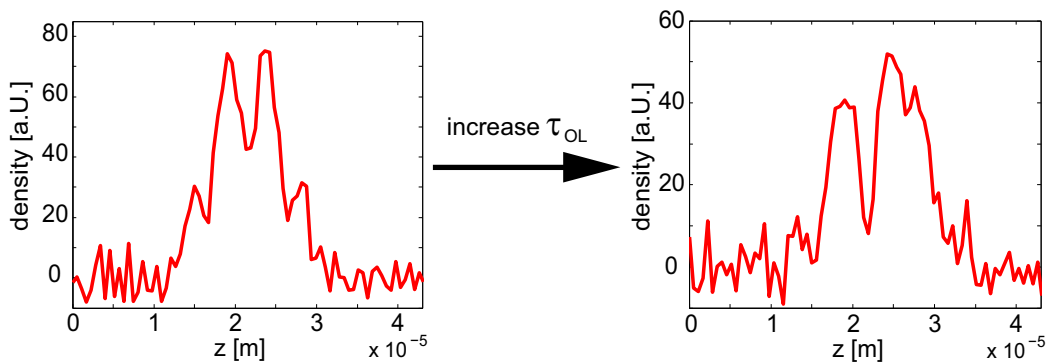}
    \caption{(Color online)
    Single shots of the longitudinal density of the BEC. Increasing the ramping-down time
    of the intensity of the optical lattice from $\tau_{OL}=2$ ms to $\tau_{OL}=5$ ms
    offers the possibility of creating a single soliton. 
    Left panel: single shot of the three
    soliton state shown in Fig. \ref{3DS_4DS}. Note that the stationary soliton in the middle appears with finite density on its minimum in the experimental images. This is due to the limited optical resolution. Right panel: single shot of a single dark soliton state. \label{1DS_3DS}
     }

\end{figure}

Our method should offer the possibility of creating a single, stationary
soliton, corresponding to the first excited state in a harmonic trap
\cite{jo}. Numerical simulations reveal that this can be achieved by
increasing the ramping-down time of the optical lattice $\tau_{OL}$
further, which decreases the kinetic energy of the collision
process. In the lowest collisional state only one interference
fringe is produced which
produces a single dark soliton.
However, for technical reasons \cite{tp}, the single (moving) dark
solitons that we created by this method (see Fig. \ref{1DS_3DS}) were fluctuating in
position from shot to shot.

The above results illustrate the possibility of experimentally generating not only
a single pair of dark solitons as in \cite{kip}, but rather of an
essentially arbitrary number of such solitons.

\section{Conclusions}

In the present work, we attempted to quantify the existence,
stability and dynamics of
multiple atomic dark solitons, by examining in detail the prototypical
cases of two- and three-dark soliton states. We provided two
complemetary viewpoints corroborating the same basic picture.
A first approach was that of considering the solitons as particles,
which interact with each other through an exponential tail-tail
interaction and are confined within a parabolic trap (appropriately
incorporating the effect of dimensionality). This particle
picture provided us with a detailed understanding of the repulsive
nature of the inter-soliton interaction and its implication on
collision-phenomena and on how it can be combined with the
restoring force of the parabolic confinement to provide for
effective stationary states of the system (i.e., of the dark
soliton ``crystal''). Within the context of these equations
of motion the normal modes of this crystal were also examined
and were associated with relative motions between the solitary
waves. The second viewpoint came from the consideration an
effective quasi-one-dimensional partial differential equation
(which incorporates appropriately the transverse confinement of
the cloud) and starting from the linear limit of the number of
atoms $N \rightarrow 0$ which has the well-known quantum harmonic
oscillator eigenfunctions and developing the multi-dark-soliton
states as natural continuations of appropriate (second- or third-
or higher-) excited modes of the linear problem. In that context,
the excitation spectrum contained the modes of the background BEC
(omitted from the particle picture), as well as the anomalous
modes pertaining to the dark-soliton quasi-particles, which were,
in turn, associated with the above mentioned normal modes.
These two approaches together with a detailed understanding of the
experimental setup of \cite{kip} provide key insights on what
types of modes can be excited in the experiment, what intrinsic
frequencies should be associated with them and, furthermore,
what types of instabilities/resonances with background excitations,
these modes can be expected to induce.

One of the future directions of the present program would be to
generalize this picture to the extent possible to the n-dark-soliton
lattice, formulating and addressing questions about the
characterization of the normal modes of such a
``dark-soliton-crystal'', as well as questions about the conditions
under which this crystal could potentially undergo phase transitions,
possibly to a state such as a ``dark-soliton-gas''. On the other
hand, another natural generalization of the present program would be
that of considering a quasi-two-dimensional analog of the waveforms,
the corresponding stability and dynamics, namely that of
multi-vortex structures and the associated particle picture. Studies
along these directions are presently in progress and will be
reported in future publications.

\vspace{5mm}

{\bf Acknowledgements}. PGK gratefully acknowledges the hospitality
of the University of Heidelberg, and the support of the Alexander
von Humboldt Foundation through a research fellowship, as well as
the support of NSF-CAREER (NSF-DMS-0349023) and of NSF-DMS-0806762.
AW acknowledges support from the Graduiertenakademie of the
University of Heidelberg. The work of DJF was partially supported by the
S.A.R.G. of the University of Athens.



\begin{thebibliography}{99}
\bibitem{tp} The long ramping down time $\tau_{OL}$ necessary to create the single soliton state results in a shallow
slope of the intensity of the optical lattice. This intensity slope cannot be locked by the feedback loop of the
position control of the optical lattice. As a result the phase difference between the two initially seperated
condensates is fluctuating. This leads to a shot to shot fluctuation of the position of the created solitons.
\end{thebibliography}
\end{document}